\documentclass[11pt]{article}
	
	\newcommand{\blind}{0}
	
    \makeatletter
    \renewcommand\section{\@startsection {section}{1}{\z@}%
                                       {-3.5ex \@plus -1ex \@minus -.2ex}%
                                       {2.3ex \@plus.2ex}%
                                       {\normalfont\fontfamily{phv}\fontsize{16}{19}\bfseries}}
    \renewcommand\subsection{\@startsection{subsection}{2}{\z@}%
                                         {-3.25ex\@plus -1ex \@minus -.2ex}%
                                         {1.5ex \@plus .2ex}%
                                         {\normalfont\fontfamily{phv}\fontsize{14}{17}\bfseries}}
    \renewcommand\subsubsection{\@startsection{subsubsection}{3}{\z@}%
                                        {-3.25ex\@plus -1ex \@minus -.2ex}%
                                         {1.5ex \@plus .2ex}%
                                         {\normalfont\normalsize\fontfamily{phv}\fontsize{14}{17}\selectfont}}
    \makeatother
	\usepackage{amsmath}
	\usepackage{graphicx}
	\usepackage{enumerate}
	\usepackage{cancel}
	\usepackage{ulem}
	\usepackage[bookmarks=false]{hyperref}
    \renewcommand{\baselinestretch}{1.5}
    \usepackage[margin=1in]{geometry}
    \usepackage{times}
    \usepackage{amsthm}
    
    \usepackage{mathtools}

    \usepackage{epsfig}
    \usepackage{amsfonts,amssymb}
    \usepackage{booktabs}
    \usepackage{tabularx,booktabs}
    \usepackage{caption}
    \usepackage{authblk}
    \usepackage{apacite}
    \usepackage{appendix}
    \usepackage{algorithm}
    \usepackage{color}
    \usepackage{geometry}
    \usepackage{setspace}
    \usepackage{natbib}
    \usepackage{amsbsy}
    \usepackage{empheq}
    \usepackage{multirow}
    \usepackage{blindtext}
    \usepackage[english]{babel}

    \usepackage{graphicx}
    \usepackage{subcaption}
    \usepackage[utf8]{inputenc}
    \usepackage[export]{adjustbox}
    \usepackage{wrapfig}
    	
\begin{document}
		
			%%%%%%%%%%%%%%%%%%%%%%%%%%%%%%%%%%%%%%%%%%%%%%%%%%%%%%%%%%%%%%%%%%%%%%%%%%%%%%
		\def\spacingset#1{\renewcommand{\baselinestretch}%
			{#1}\small\normalsize} \spacingset{1}
		%%%%%%%%%%%%%%%%%%%%%%%%%%%%%%%%%%%%%%%%%%%%%%%%%%%%%%%%%%%%%%%%%%%%%%%%%%%%%%
		
		\if0\blind
		{
			\title{\bf Hypothesis Tests with Functional Data for Surface Quality Change Detection in Surface Finishing Processes}
			\author{}
			\author{Shilan Jin$^a$, Rui Tuo$^a$, Akash Tiwari$^a$, Satish Bukkapatnam$^a$, Chantel Aracne-Ruddle$^b$, \\
			\vspace{-3mm}
			Ariel Lighty$^b$, Haley Hamza$^b$, Yu Ding$^a$$\dag$\\
			\vspace{3mm}
			$^a$Department of Industrial \& System Engineering, \\
			Texas A\&M University, College Station, Texas, USA \\
             $^b$Lawrence Livermore National Lab, Livermore, California, USA\\
						$\dag$Correspondence author, yuding@tamu.edu}
			\date{}
			\maketitle
		} \fi
		
		\if1\blind
		{

            \title{\bf \emph{IISE Transactions} \LaTeX \ Template}
			\author{Author information is purposely removed for double-blind review}
			
\bigskip
			\bigskip
			\bigskip
			\begin{center}
				{\LARGE\bf \emph{IISE Transactions} \LaTeX \ Template}
			\end{center}
			\medskip
		} \fi
		\bigskip
		
	\begin{abstract}
This work is concerned with providing a principled decision process for stopping or tool-changing in a surface finishing process.  The decision process is supposed to work for products of non-flat geometry.
The solution is based on conducting hypothesis testing on the bearing area curves from two consecutive stages of a surface finishing process. In each stage, the bearing area curves, which are in fact the nonparametric quantile curves representing the surface roughness, are extracted from surface profile measurements at a number of sampling locations on the surface of the products. The hypothesis test of these curves informs the decision makers whether there is a change in surface quality induced by the current finishing action.  When such change is detected, the current action is deemed effective and should thus continue, while when no change is detected, the effectiveness of the current action is then called into question, signaling possibly some change in the course of action. Application of the hypothesis testing-based decision procedure to both spherical and flat surfaces demonstrates the effectiveness and benefit of the proposed method and confirms its geometry-agnostic nature.
	\end{abstract}
			
	\noindent%
	{\it Keywords:} hypothesis test; functional data; inequality; mean curve; variance curve; permutation; polishing process; change detection.

	%\newpage
	\spacingset{1.5} % DON'T change the spacing!

\section{Introduction}\label{sec:introduction}
In  precision manufacturing, polishing is an inevitable post-processing step towards ensuring a nano-scale surface finish for manufactured objects when the surface roughness requirement goes beyond the capability of the manufacturing operations prior to polishing. \cite{dejule1997cmp} and \cite{frazier2014metal}, among the others, discussed extensively the necessity and impact of polishing.  A polishing process employs a polishing tool (an abrasive-embedded pad) that performs a repetitive rubbing action to remove the asperities on an object's surface in a small amount at each action.  An effective and efficient polishing process requires changing the polishing tool from time to time, as the coarse tools employed to remove large asperities cannot deliver a fine finish but polishing with fine tools would progress too slowly. The tool also steadily degrades during the process due to loading, glazing and other issues \citep{rao2015graph}. The decisions of when to change the polishing tool or to stop polishing altogether are crucial, demanding sound and consistent guidance \citep{jin2020gaussian,bukkapatnam2018planar}.

\begin{figure}[h]
    \centering
    \includegraphics[width=0.6\textwidth]{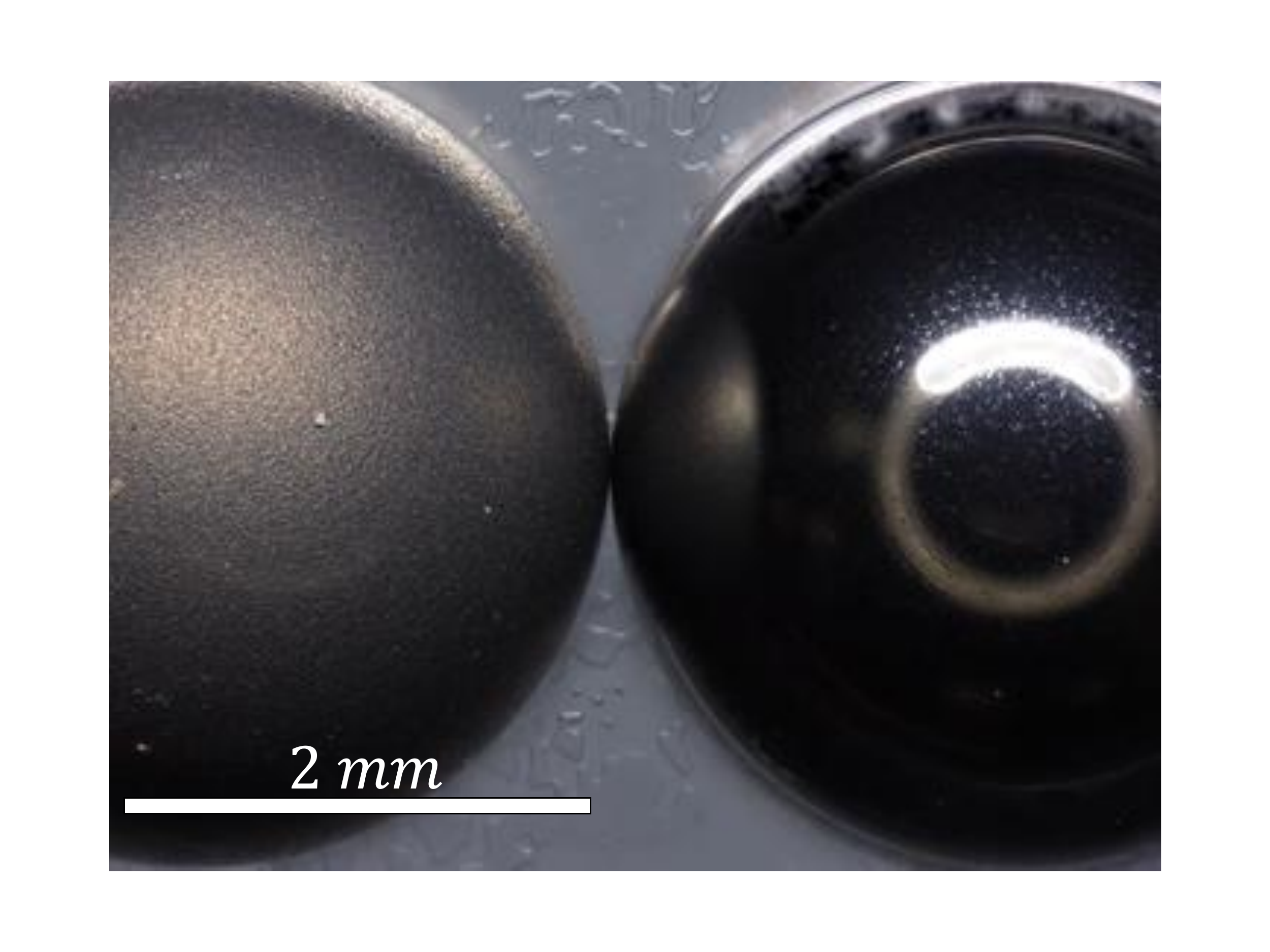}
    \caption{The silicon balls before and after being polished. }
    \label{fig:ball}
\end{figure}

In our research, we polish diamond-coated silicon balls of radius $1.688$ $mm$ to make its surface roughness to the order of $\sim 10$ $nm$ in terms of the $Ra$ value \citep{isogeometrical}. The balls are used in the physical experiments for proving the feasibility of inertial confinement fusion \citep{biener2009diamond}. During the experiments, the initial ultraviolet laser surrounding the balls is converted into soft X-rays. Bursts of the soft X-rays drive the compression of the materials in the balls to conditions similar to those found deep in the sun. The success of these experiments highly depends on the quality of the ball and its surface roughness. Figure \ref{fig:ball} presents the picture of two balls for a visual comparison of the surface before and after being polished. The left ball is unpolished and the right one is polished to about $12 nm$ in $Ra$. The picture was taken by a microscope Leica DM4000 M LED with transmitted light. The polished ball on the right reflects the ring illumination as a white circle on its surface.

The polishing process is carried out over $T$ stages, of which each stage denotes a period of time when a chosen polishing tool is applied to the surface of the polished object under the chosen conditions (like the force, rotary speed of the ball mandrel, etc). After each stage, the polishing action is paused and the surface is inspected by a microscopic measurement or imaging device, such as an optical profilometer, before the polishing action continues for the next stage. This polishing-inspection iteration repeats for every stage. What is measured after each stage affects the polishing decisions outlined earlier.

\begin{figure}[bt]
    \centering
    \includegraphics[width=0.8\textwidth]{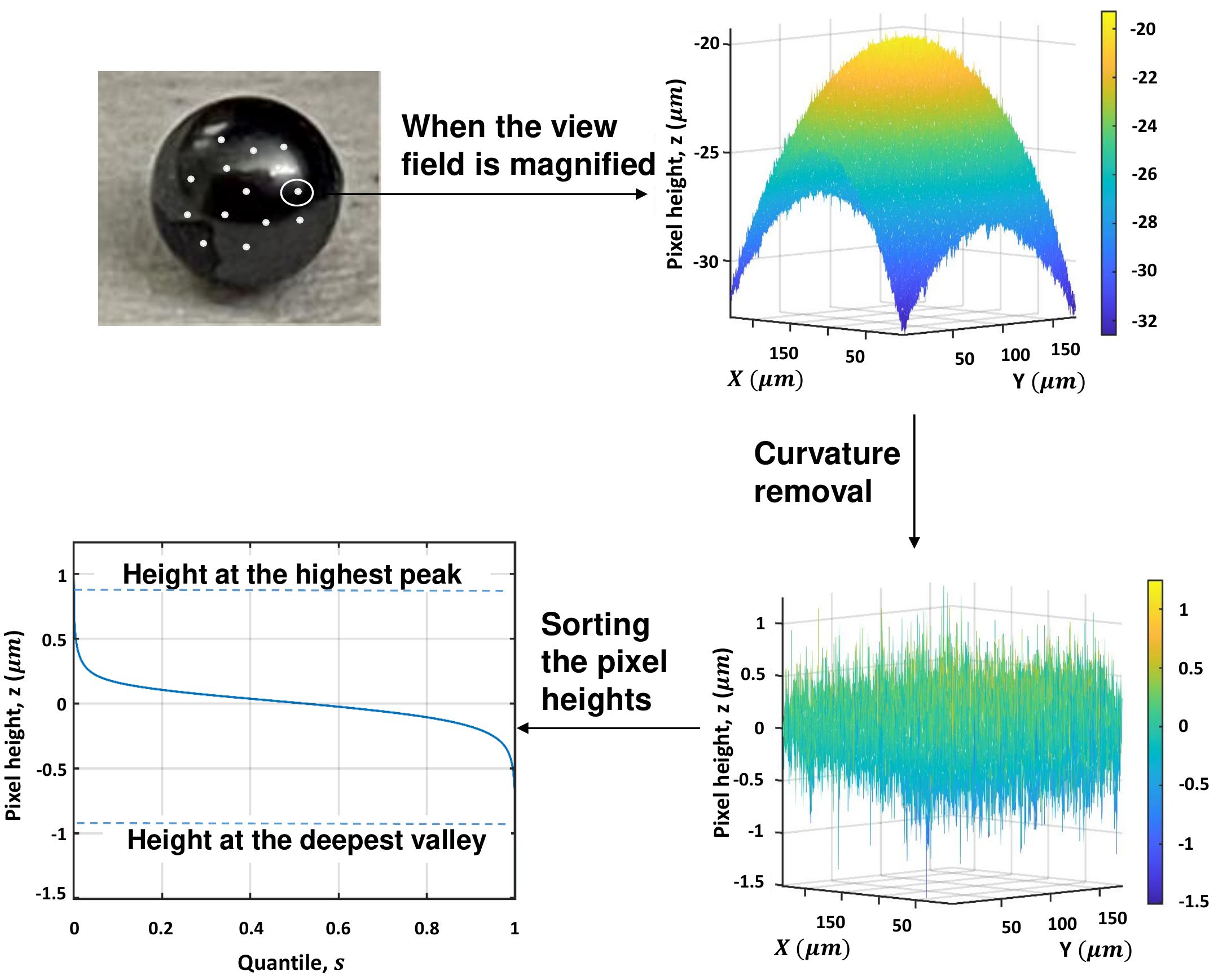}
    \caption{Top-left: the ball surface is inspected at the sample locations (white dots). Top-right and bottom-right: each location is measured by an optical profilometer which returns a matrix of pixel heights. Bottom-left: sorting the pixel heights yields a bearing area curve or the pixel-height quantile curve.}
    \label{fig:prodeg}
\end{figure}

In our specific ball polishing process, during an inspection, the profilometer measurements are taken at $M$ locations over the surface of the polished object (the white dots in Figure~\ref{fig:prodeg}, top-left panel). For each location, the profilometer returns the surface morphology covering a small area of $229.76\times 172.32$ $\mu m^2$, discretized into $640\times 480$ pixels over the $X$-$Y$ plane.  On each pixel, a height value, denoted by $z$ and in the unit of $\mu m$, is registered by the profilometer. After some data preprocessing, including curvature removal, the pixel heights are arranged into a matrix, with its row and column corresponding to the $X$ and $Y$ coordinates and its entry corresponding to the $z$ value. In the engineering practice, this matrix representation is further transformed into a one-dimensional curve, with the pixel heights sorted from the highest peak to the deepest valley; see Figure~\ref{fig:prodeg}, bottom-left panel.  Such curve is known as the \textit{bearing area curve} (BAC) in manufacturing \citep{stewart2000new} or a quantile curve in statistics. With such data arrangement, the surface quality at any stage is represented by a group of BACs (quantile curves), each of which corresponds to one of the white spots.

\begin{figure}[bt]
\centering
\begin{subfigure}{0.39\textwidth}
\includegraphics[width=\linewidth]{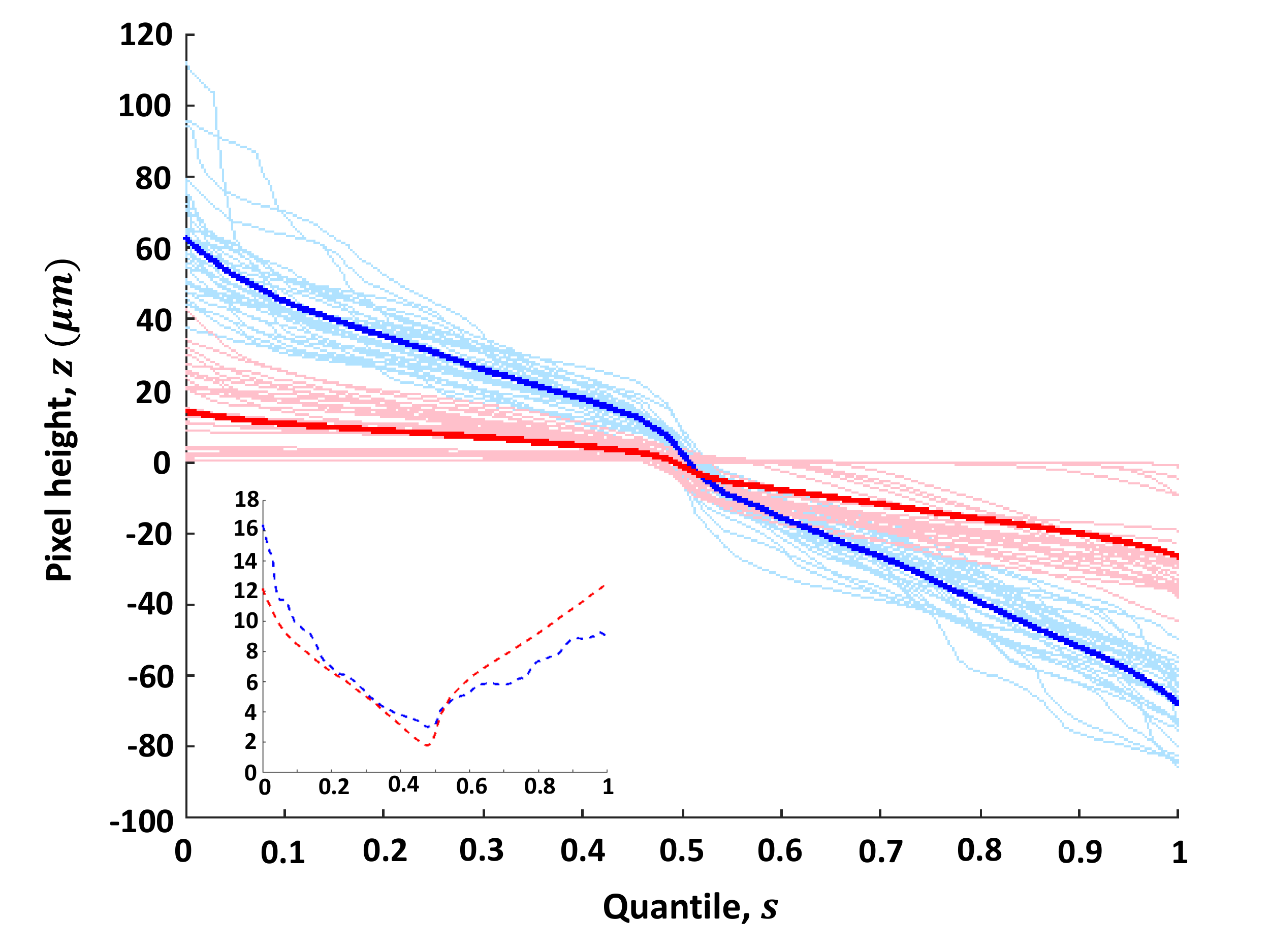}
\caption{Improvement in the tails of mean}
\end{subfigure}
\begin{subfigure}{0.19\textwidth}
\includegraphics[width=\linewidth]{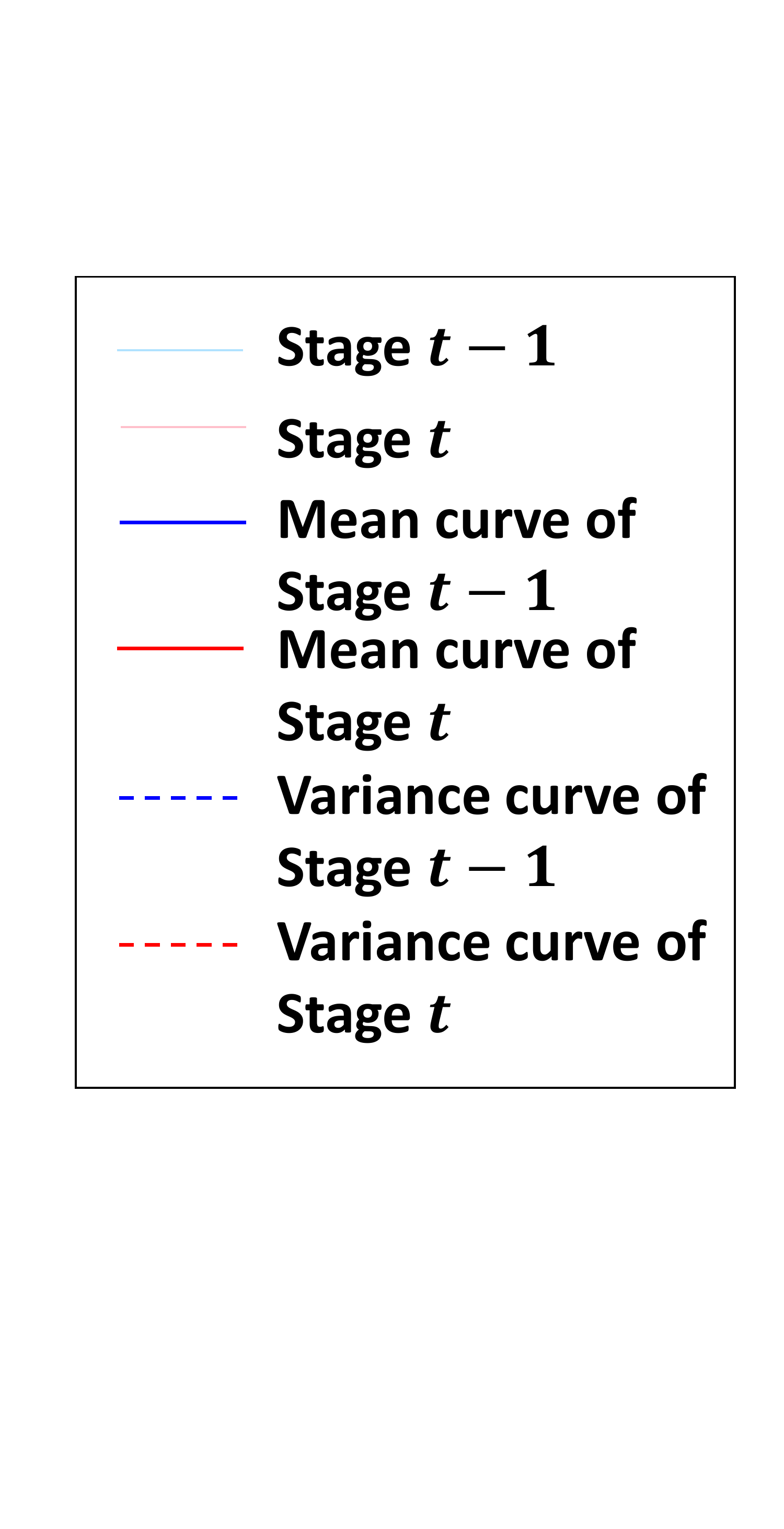}
%\caption{Legend}
\end{subfigure}
\begin{subfigure}{0.39\textwidth}
\includegraphics[width=\linewidth]{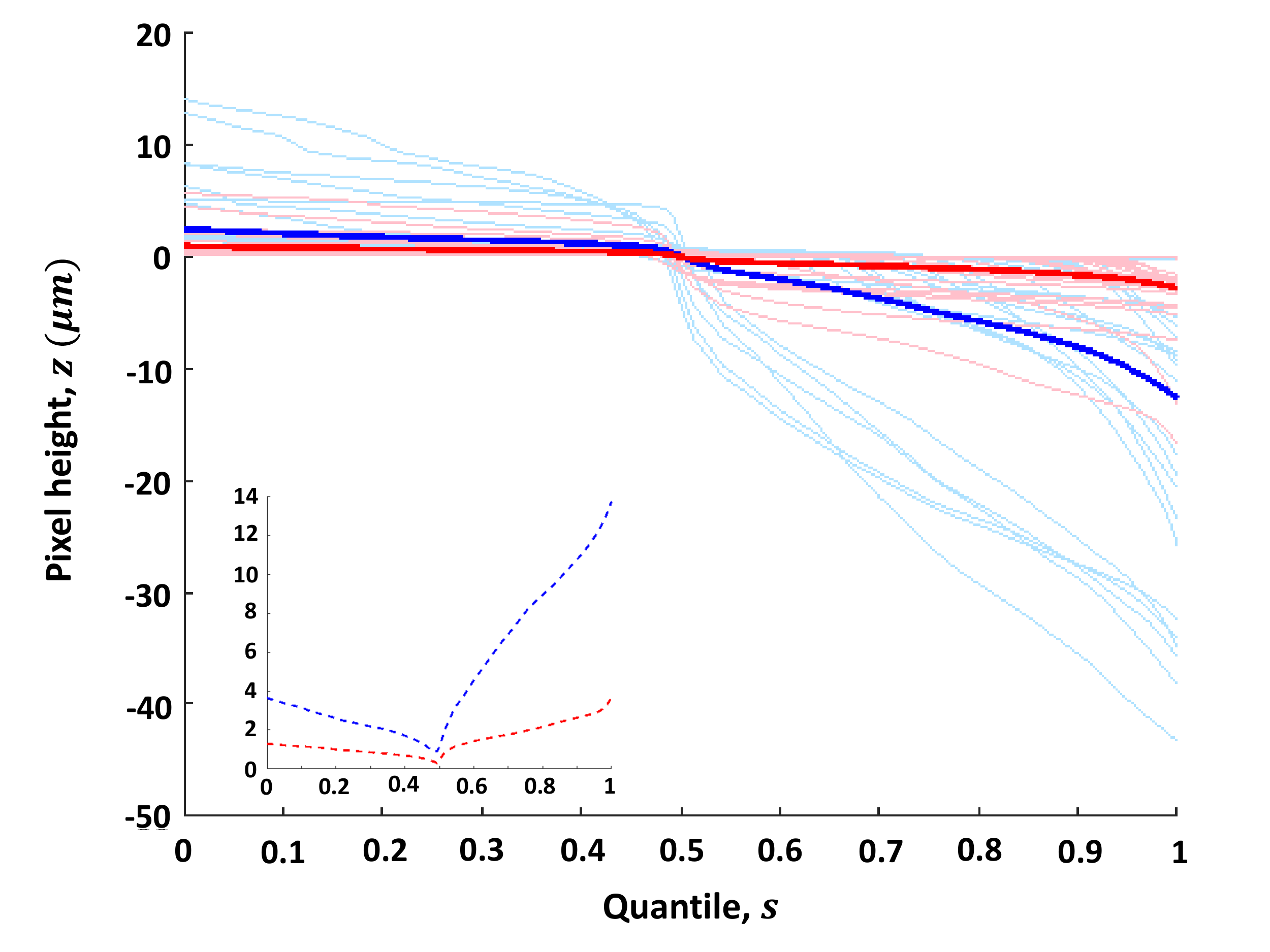}
\caption{Improvement in variance}
\end{subfigure}
\caption{Curve change across stages. In both panels, the big display shows the original curves and the mean curves inserts, whereas the small inserts show the variance curve  that is estimated by pointwise variances}.  To plot these curves, data from the first polishing experiment in \cite{jin2020gaussian} are used.
\label{fig:imp_quality}
\end{figure}

Comparing the statistical characteristics of these nonparametric BACs of two successive stages could inform us whether there is a surface quality change, caused by the polishing action between the two inspections.  When the action does make a difference to an object's surface, it most likely results in the removal of peaks and valleys, that are reflected by the upper and lower tail of the BACs respectively.  As a result, one expects to see the tails of the mean curves to be flattened (or more precisely, upper tails lowered and lower tails raised) and/or the variances to be reduced. Figure \ref{fig:imp_quality} presents two examples illustrating the observations.  Figure \ref{fig:imp_quality}(a) is at a stage where the polishing causes an obvious change in the mean curves but not so in the variance curves, whereas Figure \ref{fig:imp_quality}(b) is at a stage where the polishing improves both the mean and variance curves but the improvement in variance is more pronounced.   Should neither the mean curves nor the variance curves show much difference, it is then rather reasonable to deem the polishing action prior to the current inspection not being able to change the surface quality substantially. The detection of surface change may directly inform about polishing decisions.  When a polishing action leads to detectable surface changes, the implication is that it is effective, and consequently, the same action should be continued.  When a polishing action does not lead to detectable surface changes, it signals the need to clean or change the current tool (to a finer scale), or if the current tool is already at the finest scale, the time to stop polishing.

Motivated by this observation, we formulate the polishing decision problem to be based on the change detection in nonparametric functional curves.  Let $\mu_t(s)$ and $\sigma_t^2(s)$ be the mean and variance curve of the BACs at Stage $t$, respectively, where $s$ is the quantile variable. Our specific formulation consists of testing the following conditions:
\begin{equation}\label{eq:expmean}
\begin{split}
    &\mu_t(s)<\mu_{t-1}(s) \text{ for any } s\in[0,\tau], \quad \text{or} \\
    &\mu_t(s)>\mu_{t-1}(s) \text{ for any } s\in[1-\tau,1], \quad \text{or} \\
    &\sigma^2_t(s)<\sigma^2_{t-1}(s) \text{ for at least } 50\% \text{ of } s\in[0,1],
\end{split}
\end{equation}
where $\tau$ is a quantile cut-off for identifying the upper and lower tails.  For simplicity, we use the same $\tau$ for both tails, although it is not difficult to use two different quantile cut-offs.  In the variance test, we test the variance reduction for at least 50\% of the values because there is no guarantee that the variance is always uniformly reduced (although the right panel in Figure \ref{fig:imp_quality}(b) shows so).  Using 50\% in the test, we deem the condition true if there are more instances of reduction than otherwise.

Apparently, hypothesis testing methods on nonparameteric functional curves benefit directly our engineering decision problem; for this reason, we will provide a detailed review on nonparameteric curves testing in Section~\ref{sec:literature}. In our effort, we make a particular use of the pointwise testing method developed by \cite{cox2008pointwise}. The original method in \cite{cox2008pointwise} tests the equality of two-sample mean curves, assuming the two samples have a common covariance structure. We extend their original method to the testing of both mean and variance curves for certain sub-domains of the curves.

The merit of our research effort can be summarized as follows. The proposed statistical testing does not require the sample location information and can thus be applicable to manufactured object of  non-flat surface for which the location information is difficult to register.  This  applicability to non-flat surface, spherical surface specifically, is a major advantage in practice, considering the complexity in geometric features involved in manufactured artifacts.  The statistical testing method is able to signal subtle and detailed changes in surface roughness and appears to be a well-suited tool for the much needed polishing decisions. Because the resulting method is based on nonparametric curves, it relies on few assumptions and appears robust and easy to use. The statistical testing method, when applied to the polishing data, confirms the phenomenon of over-polishing, i.e., when the same polishing action is used for an excessively long time, it harms the surface quality rather than improving it. This finding reinforces the importance of timely decisions on tool changing or stopping. Making timely decisions leads furthermore to significant saving in time, materials and energy in a polishing process.  While we use polishing to motive the study, our proposed procedure is applicable to other surface finishing actions.  In the latter case study, we present the analysis using both polishing and lapping data.

The rest of this article is organized as follows. In Section \ref{sec:literature}, we review both the curve testing literature and the current practice of polishing decision making. Section 3 presents the proposed hypothesis testing-based method for detecting surface quality change. Section 4 applies the hypothesis testing-based method to three polishing/lapping experiments and demonstrates the impact it makes in terms of enhanced process decisions. Section 5 concludes this work.

%%%%%%%%%%%%%%%%%%%%%%%%%%%%%%%%%%%%%%%%%%%%%%%%%%%%%%%%%%%%%%%%%%%
%%%%%%%%%%%%%%%%%%%%%%%%%%%%%%%%%%%%%%%%%%%%%%%%%%%%%%%%%%%%%%%%%%%
\section{Literature Review} \label{sec:literature}
In this section, we review the literature in terms of both decision making in polishing processes and testing the difference between nonparametric functional curves.

\subsection{Decision Making in Polishing Processes}
The current industrial practice for making polishing decisions rely primarily on a simple average metrics of the surface roughness, which is the mean absolute deviation of a surface's roughness, denoted by $Sa$ \citep{isogeometrical} (or $Ra$ if it is concerned with a one-dimensional line feature). Recall the pixel height matrix of $z$ values over the $X$-$Y$ plane at a give location, as illustrated in the bottom-right plot in Figure \ref{fig:prodeg}. The $Sa$ for a location is calculated as
\begin{equation}\label{eq:Sa}
    Sa = \frac{1}{A}\iint\limits_{X, Y} \mid z(X,Y)-\overline{z}\mid dX dY,
\end{equation}
where $\overline{z}$ is the sample mean of $z(X,Y)$'s at the location and $A$ is the size of the area over which $z$ values are obtained. One $Sa$ is calculated for each location, i.e., each of the white dots in the top-left panel in Figure \ref{fig:prodeg}.  When the whole surface has $M$ sample locations, i.e., $M$ white dots, there are $M$ distinct $Sa$ values. Practitioners often use the median of the $M$ measurements of $Sa$ values, denoted by $\overline{Sa}$, to benchmark the roughness for the current stage of operations.  The popularity of median $Sa$ in practice is due to its easiness to compute/to use and its straightforward interpretability.

Recent works \citep{wang2014change, liu2017dirichlet, bukkapatnam2018planar, jin2020gaussian} showed people's awareness of a number of critical limitations of using average metric such as $Sa$ or median $Sa$ for the finishing process of precise manufacturing. \cite{jin2020gaussian} provided specific examples where a set of similar media $Sa$ values correspond to surfaces with rather different roughness features. \cite{jin2020gaussian} further proposed a new decision making criterion for polishing by modeling the surface roughness at each stage through a Gaussian process (GP). The surface of the polished object is treated like a landscape, so that its analogy to spatial statistics is invoked, explaining why the GP model is used.  \cite{jin2020gaussian} demonstrated that the scale parameter in the resulting GP model is sensitive to changes in the surface roughness and consequently devised a decision procedure based on that.  While \cite{jin2020gaussian} took an important step forward in introducing sophisticated statistical modeling to decision making in polishing, there are still two limitations in their method.

The first is that the use of a single scale parameter in the GP model as the representation of the surface roughness still compresses the detailed surface quality information and may lead to information loss in subsequent decision processes. The second limitation is that the applicability of the resulting GP-based method is limited to flat (or nearly flat) surfaces, due to the use of a squared-exponential covariance structure in the GP model. When the surface has a strong curvature, like the ball-surface polishing here, or in some other cases where complicated geometric features are involved, either the covariance function in the GP model must have a fundamental redesign (to account for the effect of geodesic distances), so that the GP-based decision rules could be extended to non-flat surfaces, or a new  geometry-friendly decision process  needs to be designed.

Accommodating the aforementioned two requirements, we choose to propose a new geometry-friendly  decision process,  to be based on hypothesis testing of nonparametric quantile curves (the BACs) associated with two consecutive surface finishing stages.

\subsection{Hypothesis Tests with Nonparametric Functional Data}
Hypothesis tests of nonparametric functions entail the mean function test and the covariance structure test. In terms of the mean function test, for example, \cite{hall1990bootstrap} studied a test statistic produced by the difference of the estimations of the two functions, and proposed to use bootstrap to estimate the distribution of the test statistic. \cite{king1991testing} modified \cite{hall1990bootstrap}'s test statistic, so that an asymptotic distribution can be attained under the Gaussianality assumption. \cite{kulasekera1995comparison}, \cite{munk1998nonparametric} and \cite{neumeyer2003nonparametric} investigated their respective test statistics. Chapters 5 and 9 in \cite{zhang2013analysis} tested the mean functions of the samples of two curves through testing the $L_2$ difference between the two mean functions.

In terms of variance tests, the common technique is to use Karhunen-Lo\`eve expansion to approximate the continuous covariance operators with a finite set of functional principal components (FPCs), and test the equality of each component between two samples. \cite{benko2009common} tested the equality tests of both mean functions and covariance structures. They constructed the test statistics that were the distance of the two-sample FPCs and used bootstrap to estimate the distributions of the test statistics. Some other works were devoted to the testing of the covariance structures.  For instance, \cite{panaretos2010second} constructed a test statistic to test the FPCs assuming that the functional data follows Gaussian processes. \cite{fremdt2013testing} tested the same type of problem as \cite{panaretos2010second}'s but relaxed the Gaussian process assumption.  Chapter 10 of \cite{zhang2013analysis} discussed the test of two covariance functions by defining an $L_2$-based test statistic, measuring the difference between two covariance matrices.

The above-referenced works produce outcomes for a so-called \emph{global} test, meaning that they give a binary answer concerning whether the two sets of functional curves are the same or not.  But they did not identify where the differences may lie nor do these methods work for a subset of the input domain.  Recall that the hypothesis tests we envision for the polishing process need to test for the tail portions or a subdomain (at least $50\%$), rendering these method not directly applicable to our engineering problem.

To the best of our knowledge, there are two studies that can do a \emph{local} test, i.e., the test can be done to a subdomain, as the nature of our problem requires. One is \cite{cox2008pointwise} and the other is \cite{prakash2020gaussian}. \cite{cox2008pointwise} studied the testing problem when there are curve replicates and the input locations where the curves are sampled are the same between the two groups of curves. \cite{prakash2020gaussian} took advantage of the Bayesian posterior covariance to build a confidence band for the mean function difference. Their method is applicable to the circumstance where there is no curve replicate (a single curve in each test group) and the input locations are different.

Our problem setting matches with that of \cite{cox2008pointwise} much better.  That is why we choose to follow \cite{cox2008pointwise}. But \cite{cox2008pointwise} only presented a mean function test over the whole domain.  To solve our problem, we need to introduce a mean function test for the tail portions and a variance test. Further we need to combine all the tests for devising a unified decision rule for making polishing decisions.

\section{Tests and the Detection Rule} \label{sec:tests}
In this section, we discuss how we expand the hypothesis test with functional data proposed by \cite{cox2008pointwise}, so that the testing method can be applied to the inequality test of two-sample mean functions and that of two-sample variance functions. The new test provides the basis for designing a rule for detecting the surface quality change.

 The basic idea of \cite{cox2008pointwise} is to approximate the curve test with multiple pointwise tests of points on the curve. Considering the multiple tests as a family, to control the familywise error rate, they propose to use \cite{westfall1993resampling}'s permutation method, as the family of tests are highly correlated when the points are close to each other. We extend the method of  \cite{cox2008pointwise} and establish test statistics corresponding to different hypothesis test statements demanded in polishing decision processes.

\subsection{Hypothesis Tests}

Two groups of curves are represented with $z_{tj}(x) $, where $t=1,2$ is the group index and $j=1,\cdots,J_t$ is the curve index. Without loss of generality, let $x\in[0, 1]$, meaning the input value is normalized to the range of $[0, 1]$. We want to test for the existence of an inequality between $\mu_1(x)$ and $\mu_2(x)$, that is the population mean of two groups of curves, and for the existence of an inequality between $\sigma^2_1(x)$ and $\sigma^2_2(x)$, that is the population variance of two groups of curves, for a certain subdomain of the curves.  Denote by $\varphi(x)$ either $\mu(x)$ or $\sigma^2(x)$. The subdomain is denoted by $A \subset [0,1]$. There are multiple options for this subdomain $A$; for instance, it could be all $x\in[0,1]$, at least $50\%$ of $x\in[0,1]$, or at least one $x\in[0,1]$.

Using the notations just introduced, the general form of the hypothesis test of functional inequality can be expressed as
\begin{equation}\label{eq:gen}
%\begin{split}
    H_0^{\varphi}: \varphi_1(x)=\varphi_2(x) \text{ for all } x\in [0,1]
    \text{ against } H_1^{\varphi}: \varphi_1(x)>\varphi_2(x)\text{ for } x\in A.
 %   \end{split}
\end{equation}
A few specific hypotheses are given in the following:

\begin{enumerate}
    \item Test on the mean function, regardless of the relationship between $\sigma_1^2(x)$ and $\sigma_2^2(x)$:
\begin{equation}\label{eq:genmean}
    \begin{split}
        &H_0^{\mu}: \mu_{1}(x)=\mu_2(x) \text{ for all } x\in[0,1]\\
    \text{  against  }
    &H_1^{\mu}: \mu_{1}(x)>\mu_2(x) \text{ for all } x\in[0,1].
    \end{split}
\end{equation}In this example, $\varphi(x)=\mu(x)$ and $A=\{x: \text{ all } x\in[0,1]\}$.

\item Test on the variance function, regardless of the relationship between $\mu_1(x)$ and $\mu_2(x)$:
\begin{equation}\label{eq:genvar}
\begin{split}
    &H_0^{\sigma^2}: \sigma_{1}^2(x)=\sigma_{2}^2(x) \text{ for all }x\in[0,1]\\
    \text{ against } &H_1^{\sigma^2}: \sigma_{1}^2(x)>\sigma_{2}^2(x) \text{ for at least } 50\% \text{ of }x\in[0,1],
\end{split}
\end{equation} In this example, $\varphi(x)=\sigma^2(x)$ and $A=\{\text{at least }50\% \text{ of }x\in[0,1]\}$.
\end{enumerate}

\cite{cox2008pointwise} presented a method valid for testing the equality of two mean curves, assuming that they have the identical covariance structure. Their hypothesis test is similar to the first example, i.e., in \eqref{eq:genmean}, except that the alternative hypothesis is stated as $H_1: \mu_1(x)\neq \mu_2(x)$ for at least one $x\in[0,1]$. Their method discretizes the continuous domain of the curve into a dense grid of points and performs univariate, pointwise $t$-tests for two-sample mean comparisons. Cox and Lee showed that despite the pointwise tests, their method is in fact a functional comparison. They employed the Westfall-Young permutation-based randomization procedure to control for the familywise error rate. Our test procedures borrow the ideas in \cite{cox2008pointwise} but need to expand the test statistics  and adjust the permutation steps (discussed in Section \ref{sec:fwer}).

\subsubsection{Pointwise Test}\label{sec:pointwise}
Because \cite{cox2008pointwise} employed pointwise tests, they discretize the continuous domain of $x$ to a grid of  points, and let us denote the discretized domain by $D_{m} \subset [0,1]$. Accordingly, the discrete counterpart of $A\subset [0,1]$ is denoted by $A_m\subset D_m$. For instance, the discrete counter part of $A(x)=\{\text{at least }50\% \text{ of }x\in[0,1]\}$ is $A_m(x)=\{\text{at least }50\% \text{ of }x\in D_m\subset[0,1]\}$.  Using the discrete set notation, the pointwise testing problems can be expressed as:
\begin{equation}\label{eq:disgen}
\begin{split}
    & H_0^{\varphi}(x): \varphi_1(x)=\varphi_2(x) \text{ for all }x\in D_m\\
    \text{ against } & H_1^{\varphi}(x): \varphi_1(x)>\varphi_2(x) \text{ for all }  x\in A_m,
\end{split}
\end{equation}

For the hypotheses in (\ref{eq:disgen}), individual univariate hypothesis tests are conducted to compare the two groups of curves evaluated at each input point $x$. The individual test can be either a one-sided $t$-test, for testing the inequality of mean, i.e., when $\varphi=\mu$, or a $F$-test, for testing the inequality of variance, i.e., when $\varphi=\sigma^2$. Each of the univariate two-sample tests computes a $p$-value, denoted by $p(x)$, at each grid point $x$. Let us further denote the $p$-values computed from the original data samples by $p^o(x)$, where the superscript $o$ indicates the \textit{original $p$-values} or \textit{observed $p$-values}. This superscript notation is introduced to differentiate the original $p$-values from the $p$-values generated from the re-sampled data. Recall that a smaller $p$-value presents stronger evidence in favor of $H_1(x): \varphi_1(x)>\varphi_2(x)$.

\subsubsection{Control Familywise Error Rate} \label{sec:fwer}
Using the pointwise tests, the comparison in continuum is approximated by a multiple comparison problem.  We need to find a test statistic such that the family of the pointwise tests can be under the control of a prescribed significance level of $\alpha$.

Analogous to the type I error, the familywise error rate (FWER) is the probability of the false rejection of the complete null hypothesis that consists of multiple comparisons. That is, given $\alpha$,
\begin{equation} \label{eq:defoferror}
    Pr\{H_0^\varphi \text{ is rejected } | {H_0^\varphi} \text{ is true }\} \leq \alpha.
\end{equation}

 For a univariate test, the p-value is uniformly distributed under the null hypothesis  \citep{klammer2009statistical}. That is $Pr\{P\leq\alpha\}\leq\alpha$, where $P$ denotes the p-value as a random variable. With the family of null hypotheses, we want the joint probability of rejecting all the null hypotheses, given that they are true, is less than $\alpha$. The conventional Bonferroni correction does not perform well for multiple comparisons involving functional data, because the events of rejection of these individual null hypotheses are highly correlated when the input locations ($x$'s), with which the hypotheses are associated, are close. But can one still control the familywise error rate considering the inherent correlations in the functional data? The short answer is yes. Westfall-Young's permutation-based procedure \citep{westfall1993resampling} provides the ability to do so. Before diving into the specific procedure, we need to understand what we want to control, i.e., the test statistics.

\textbf{Test Statistics.}
The choice of the test statistic for the purpose of FWER control depends on the subdomain of interest of $A_m$:
\begin{itemize}
    \item If $A_m=\{\text{at least one } x \in D_m\}$, the test statistic has been studied by \cite{cox2008pointwise}, which is $\text{min}P$, the minimal $p$-value among those computed from the family of individual tests. This is to say, as long as one $x$ exists, such that $\varphi_1(x)>\varphi_2(x)$ does not occur by chance, we accept $H_1^\varphi(x)$. Equation (\ref{eq:defoferror}) then becomes
    \begin{equation}\label{eq:minp}
        Pr\{\text{min}P\leq\alpha\}\leq\alpha.
    \end{equation}
    We further denote the realization of $\text{min}P$ by $\text{min}p$.

    \item If $A_m=\{\text{all }x\in D_m\}$ , the test statistic is $\text{max}P$, the maximal pointwise $p$-value. We control the probability of most probably event, that is, the probability of the individual test, $\varphi_1(x)>\varphi_2(x)$, evaluated at the $x$ that returns the maximal $p$-value among all $x\in A_m$. Should this individual test be under the significance level of $\alpha$,  the $p$-values of every other event will be controlled as well. Equation (\ref{eq:defoferror}) then becomes
    \begin{equation}\label{eq:maxp}
        Pr\{\text{max}P\leq\alpha\}\leq\alpha.
    \end{equation}
    We further denote the realization of $\text{max}P$ by $\text{max}p$.

    \item If $A_m=\{\text{at least }50\% \text{ of }x\in D_m\}$, the test statistic is $\text{med}P$, the median of the pointwise $p$-values.   We control the median pointwise $p$-value to be smaller than or equal to the significance level of $\alpha$, so that the events of $\varphi_1(x)>\varphi_2(x)$ associated with the $p$-values smaller than the median $p$-value will be controlled as well. Equation (\ref{eq:defoferror}) then becomes
    \begin{equation}\label{eq:medp}
        Pr\{\text{med}P\leq\alpha\}\leq\alpha.
    \end{equation}
    We further denote the realization of $\text{med}P$ by $\text{med}p$.
\end{itemize}

\textbf{Procedure of the Test.}
For generating the distribution of $\text{min}P$, or $\text{max}P$, or $\text{med}P$, \cite{westfall1993resampling} presented a permutation-based procedure.  We use $\text{min}P$ as an example to illustrate how to apply Westfall-Young's procedure to generate its distribution. For generating the distribution of $\text{max}P$ or $\text{med}P$, simply change the boldface $\textbf{min}$ in the following with either $\text{max}$ or $\text{med}$, respectively.

Recall that given the two samples of curves, we calculate the pointwise $p$-values under the null hypothesis and record the minimal $p$-value as $p_\textbf{min}^o$.  We then randomly permute the curves between the two samples $N$ times. At each iteration, the two groups of resampling curves are generated. With the resampling data, we perform the pointwise tests and report the $p$-value of each individual test, denoted by $p_l^s(x)$, $l=1,\cdots,N$, where the superscript $s$ indicates \textit{simulated} $p$-values.  Then, calculate $\textbf{min}_{x\in A_m} (p_l^s(x))$  and denote it by $\textbf{min} p_l^s$.  Generate the empirical distribution of $\textbf{min}P$ using $\{\textbf{min} p_l^s$, $l=1,\cdots,N\}$. The corrected $p$-value with the FWER controlled for is $Pr\{\textbf{min}P < p_{\textbf{min}}^o\}$, which indicates how extreme the observed familywise minimal $p$-value is.  This corrected $p$-value can be empirically estimated by using the empirical distribution of $\textbf{min}P$ obtained through the permutation procedure. Specifically, the estimate of the corrected $p$-value is through locating $p_{\textbf{min}}^o = \textbf{min}_{x\in A_m} p^o(x)$ on the empirical distribution of $\textbf{min}P$. The detailed steps of the test procedure is presented in Algorithm \ref{alg}.

\begin{algorithm}[h]
\caption{Estimate the p-value of the functional curve test}
\label{alg}
For \textit{stat} = min, max or med,
\begin{itemize}
\item[] \textbf{Step 1.} Perform univariate one-sided $t$ tests for testing mean (or $F$ tests for testing variance) on $x_k\in A_m$, $k=1,\cdots,m$. Compute the pointwise $p$-values, $p(x_k)$ and  then $p_{\textit{stat}}=\textit{stat}\{p(x_k):k=1,\cdots,m\}$.
\item[] \textbf{Step 2.} Let $p^o_{\textit{stat}}\leftarrow p_{\textit{stat}}$.
\item[] \textbf{Step 3.} For $l=1,\cdots, N$, randomly permute the group label $t\in\{1,2\}$ in the data $\{z_{jt}(x_k): j=1,\cdots,J_t, k=1,\cdots,m\}$. Repeat \textbf{Step 1}. Record $p^{(l)}=p_{\textit{stat}}$.
\item[] \textbf{Step 4.} Find $l_0$, such that $p^{(1)}\leq\cdots\leq p^{(l_0)}\leq\cdots\leq p^{(N)}$, where $p^{(l_0)}=p_{\textit{stat}}^o$.
\item[] \textbf{Step 5.} The corrected $p$-value then is $\frac{l_0}{N}$.
\end{itemize}
\end{algorithm}

We permute the whole curves between the two groups to preserve the inherent correlation among the points within a curve.  This treatment is different from the practice in \cite{cox2008pointwise}, in which the points are allowed to permute between the two groups.  \cite{cox2008pointwise}'s permutation does not cause any problem, as they assume a common covariance structure of the two groups of the curves, meaning that two samples of discrete points follow the same joint distribution. In our engineering decision process, such assumption cannot be guaranteed. Permuting the curves as a whole relaxes such requirements and doing so is also consistent with engineering practice, as an assembly of curve segments from different stages does not have a valid engineering meaning.

\subsection{Hypothesis Tests Based Detection Rule}\label{sec:rule}
As discussed in Section \ref{sec:introduction}, when a polishing action improves the surface quality, one expects to notice either the mean curves, especially its tail portions, are flattened, or the variance of the curves is reduced, or both. In other words, if any or all of the inequalities in Equation \ref{eq:expmean} holds, it signals an improvement of the surface quality, suggesting that the polishing action is effective.  Understandably, we would like to test the following three hypotheses on functional curves:

\begin{equation} \label{eq:h1}
\begin{cases}
    H_0^{\mu_{up}}:
    \mu_{t-1}(s)=\mu_{t}(s) \text{ for all } s\in[0,\tau] \\
    H_1^{\mu_{up}}:
    \mu_{t-1}(s)>\mu_{t}(s) \text{ for all } s\in[0,\tau] \;\;\;\;\;\;\;\;\;\;\;\;\;\;\;\;\;\;
\end{cases}
\end{equation}
\begin{equation}\label{eq:h2}
\begin{cases}
    H_0^{\mu_{lo}}:
    \mu_{t-1}(s)=\mu_{t}(s) \text{ for all } s\in[1-\tau,1] \\
    H_1^{\mu_{lo}}:
    \mu_{t-1}(s)<\mu_{t}(s) \text{ for all } s\in[1-\tau,1] \;\;\;\;\;\;\;\;\;\;\;\;\;
\end{cases}
\end{equation}
\begin{equation}\label{eq:h3}
\begin{cases}
     H_0^{\sigma^2}:
    \sigma_{t-1}^2(s)=\sigma_{t}^2(s) \text{ for all }  s\in [0,1] \\
    H_1^{\sigma^2}:
   \sigma_{t-1}^2(s)>\sigma_{t}^2(s) \text{ for at least } 50\% \text{ of } s\in [0,1].
\end{cases}
\end{equation}
%%%%%%%%%%%%%%%%%%%%%

The superscripts, $\mu_{up}$ and $\mu_{lo}$, indicate that the respective hypothesis is tested for the upper tail or lower tail of the mean functions, respectively, and the superscript, $\sigma^2$, indicates the variance function test. These hypothesis tests have clear physical interpretation. The alternative hypothesis of Equation (\ref{eq:h1}) suggests that the peaks are being flattened; the alternative hypothesis of Equation \eqref{eq:h2} suggests that the valleys are being filled; and the alternative hypothesis of Equation (\ref{eq:h3}) means that the surface is getting more even. A null hypothesis is rejected when there is strong evidence against it. The strength of the evidence is quantified by the $p$-value; the smaller, the stronger.

To perform the pointwise tests, we generate a grid of evaluation points $D_m=\{s_1, \cdots, s_m\} \subset [0,1]$. We evaluate every function $z_{tj}(s)$ at these finite points in $D_m$, yielding the data vectors ${z}_{tj}=(z_{tj}(s_1),\cdots,z_{tj}(s_m))'$. Define $D_{u} = \{s_k, k=1,\cdots,u: s_{
k}\leq \tau, s_k\in D_m\}$ and $D_{l} = \{s_k,k=m-l,\cdots,m: s_{k}\geq 1-\tau, s_k\in D_m\}$. Upon discretizing the continuum of comparisons, the three functional hypotheses are approximated with three families of individual univariate hypotheses:

\begin{equation}\label{eq:dis_h1}
\begin{cases}
    H_0^{\mu_{up}}(s): \mu_{t-1}(s)=\mu_{t}(s) \text{ for all } s\in D_{u} \subset [0,\tau], \\
    H_1^{\mu_{up}}(s): \mu_{t-1}(s)>\mu_{t}(s) \text{ for all } s\in D_{u} \subset [0,\tau].\;\;\;\;\;\;\;\;\;\;\;\;\;\;\;\;\;\;
  \end{cases}
\end{equation}
\begin{equation}\label{eq:dis_h2}
    \begin{cases}
       H_0^{\mu_{lo}}(s):  \mu_{t-1}(s)=\mu_{t}(s) \text{ for all } s\in D_{l} \subset [1-\tau,1],\\
       H_1^{\mu_{lo}}(s):  \mu_{t-1}(s)<\mu_{t}(s) \text{ for all } s\in D_{l} \subset [1-\tau,1].\;\;\;\;\;\;\quad\quad\;
    \end{cases}
\end{equation}
\begin{equation}\label{eq:dis_h3}
    \begin{cases}
       H_0^{\sigma^2}(s): \sigma_{t-1}^2(s)=\sigma_{t}^2(s) \text{ for all }  s\in  D_{m} \subset [0,1],\\
       H_1^{\sigma^2}(s): \sigma_{t-1}^2(s)>\sigma_{t}^2(s) \text{ for at least } 50\% \text{ of } s\in D_{m}\subset [0,1].
    \end{cases}
\end{equation}

In the tests, $\mu_t(s)$ and $\sigma^2_t(s)$ are estimated by their sample counterparts, i.e., sample mean curves, $\overline{z}_t(s)$, and sample variance curve, $\varsigma_t^2(s)$, such that
\begin{equation}
\overline{z}_t(s)=\frac{1}{J_t}\sum_{j=1}^{J_t}z_{tj}(s) \quad \text{and} \quad \varsigma_t^2(s) = \frac{1}{J_t-1}\sum_{j=1}^{J_t}(z_{tj}(s)-\overline{z}_t(s))^2.
\end{equation}
 We control the familywise error rate of the hypothesis in Equations (\ref{eq:dis_h1}) and (\ref{eq:dis_h2}) using the test statistic $\text{max}P$ as in Equation (\ref{eq:maxp}), and control the familywise error rate of the hypothesis in Equation (\ref{eq:dis_h3}) using the test statistic $\text{med}P$ as in Equation (\ref{eq:medp}).  The corresponding testing procedure is described in Section \ref{sec:fwer}.  

Combining the three tests, our rule for surface change detection is---\textit{if the null hypothesis of any of the three sub-families is rejected, it signals that the current polishing action is effective and should be continued; otherwise, the polishing action did not make statistically significant change to the surface under polishing.} To control the type-I error of this final detection rule, we invoke the Bonferroni's method for multiple comparisons.  The three families of tests, as in Equations (\ref{eq:dis_h1})-(\ref{eq:dis_h3}), can be considered independent with each other, so that Bonferroni's method sets the type-I error rate for each of the test families to be less than or equal to $\alpha/3$. Within each family, these pointwise tests are highly correlated, so that the respective FWER is controlled through the Westfall-Young's procedure, as outlined in Algorithm \ref{alg}.

Please note that the above decision rule is for detection of a surface change.  Once a surface change is not detected, what and how engineers should react to it needs further consideration.  Generally speaking, there are two options: (a) cleaning the polishing tool or (b) changing to a finer tool.  The second option also leads to a stopping decision of the overall manufacturing post-process, if the finest polishing tool has already been used.

Between the choice of cleaning the tool and changing to a finer tool, the decision is usually not hard.  It does not cost much to clean the tool and use the clean tool to polish a little further.  When there is still no improvement in the surface quality, it is then clear that a new tool is needed, or if no finer tool is there to be used, the polishing process should be naturally stopped at that point.

\section{Physical Experiments}\label{sec:phyexp}

We apply the proposed hypothesis test based detection method to the polishing/lapping process of the spherical surface of silicon beads, introduced in Section  \ref{sec:introduction}. To achieve its nano-scale finish, the coated spherical bead is passed through the tip truncating (TT) process, the lapping process and the polishing process; the tools are refined as the beads go through these different processes. Each type of processes is divided into several stages so the surface roughness at the corresponding intermediate stages can be measured. See Figure \ref{fig:prodeg} for the illustration of the measurement and arrangement of the surface roughness data at a process stage.

\begin{figure}[h!]
     \centering
     \begin{subfigure}[b]{0.47\textwidth}
         \centering
         \includegraphics[width=\textwidth]{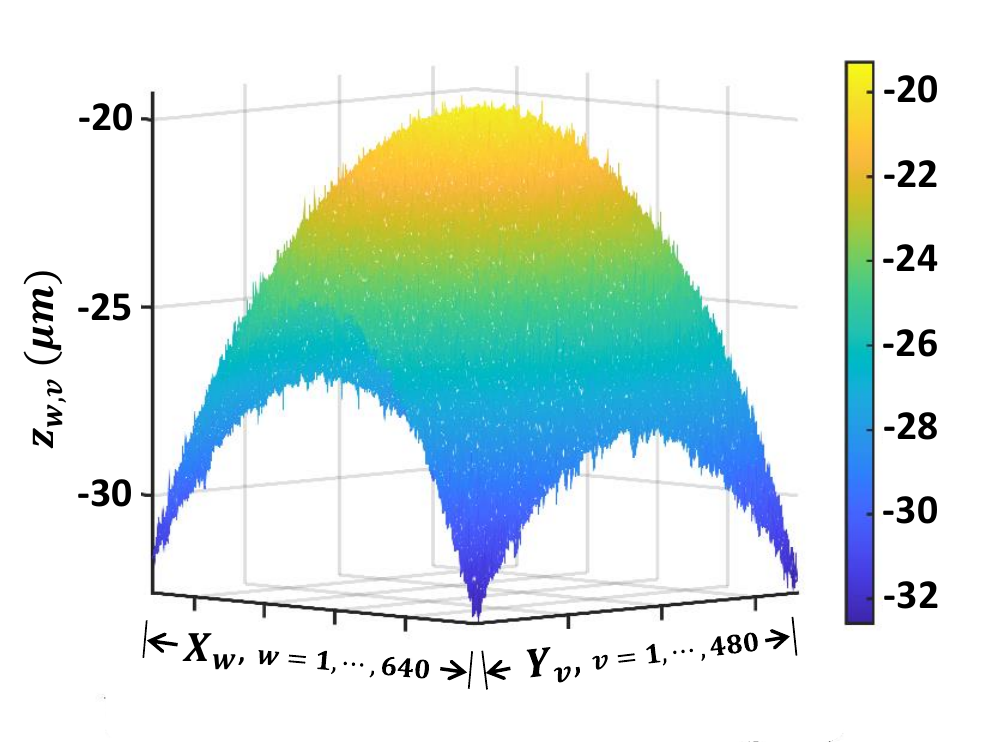}
         \caption{}
        %  \label{}
     \end{subfigure}
     %\hfill
     \hspace{15mm}
     \begin{subfigure}[b]{0.4\textwidth}
         \centering
         \includegraphics[width=\textwidth]{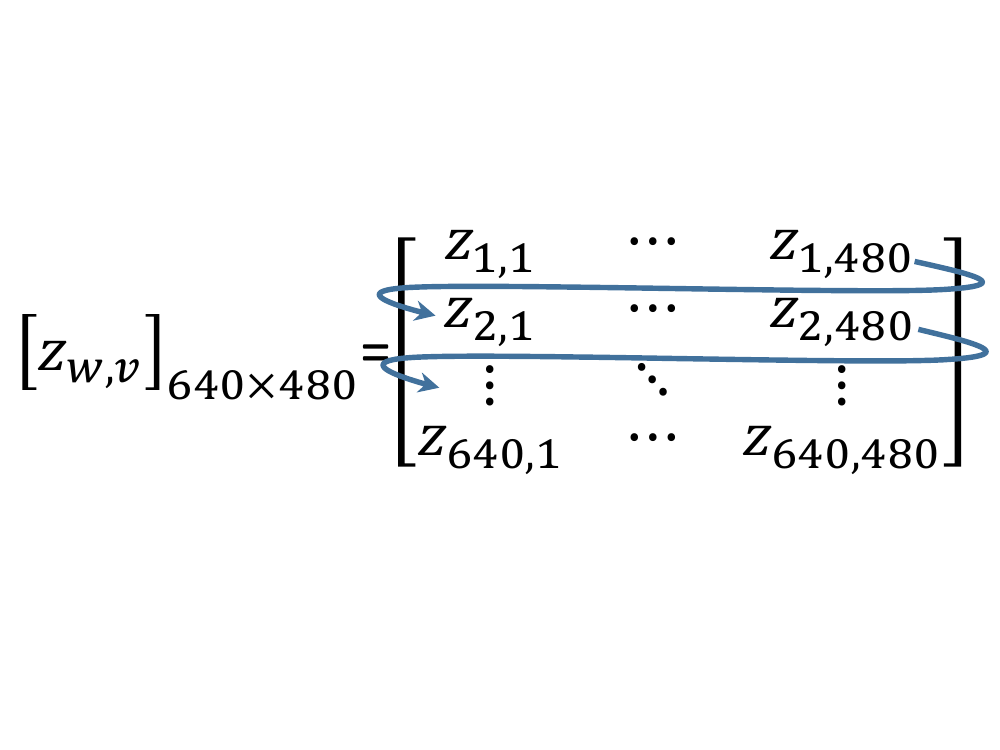}
         \caption{}
        %  \label{}
     \end{subfigure}
     \caption{(a) Pixel height matrix at a given location. The optical profilometer scans an area of $229.76\times 172.32 \,\, \mu m^2$ and save the pixel heights into a matrix $Z$ of $640\times480$ pixels.  (b) The matrix $Z$ is converted to a vector of $z$. }
     \label{fig:data}
\end{figure}

Figure \ref{fig:data} (a) demonstrates the data configuration of the surface roughness in a matrix of $Z$, where the pixel height is represented by $[z_{w,v}]$ and $w, v$ are the pixel position indices along the two directions, respectively. The $X$ and $Y$ coordinates are denoted by $X_w$ and $Y_v$, respectively.  The pixels are spread evenly over the whole surface, with a between-pixel distance of $0.359$ $\mu m$ along the $X$ coordinate and $0.369$ $\mu m$ along the $Y$ coordinate.

\subsection{Pixel Height Calibration}
\begin{figure}[h!]
     \centering
     \begin{subfigure}[b]{0.49\textwidth}
         \centering
         \includegraphics[width=\textwidth]{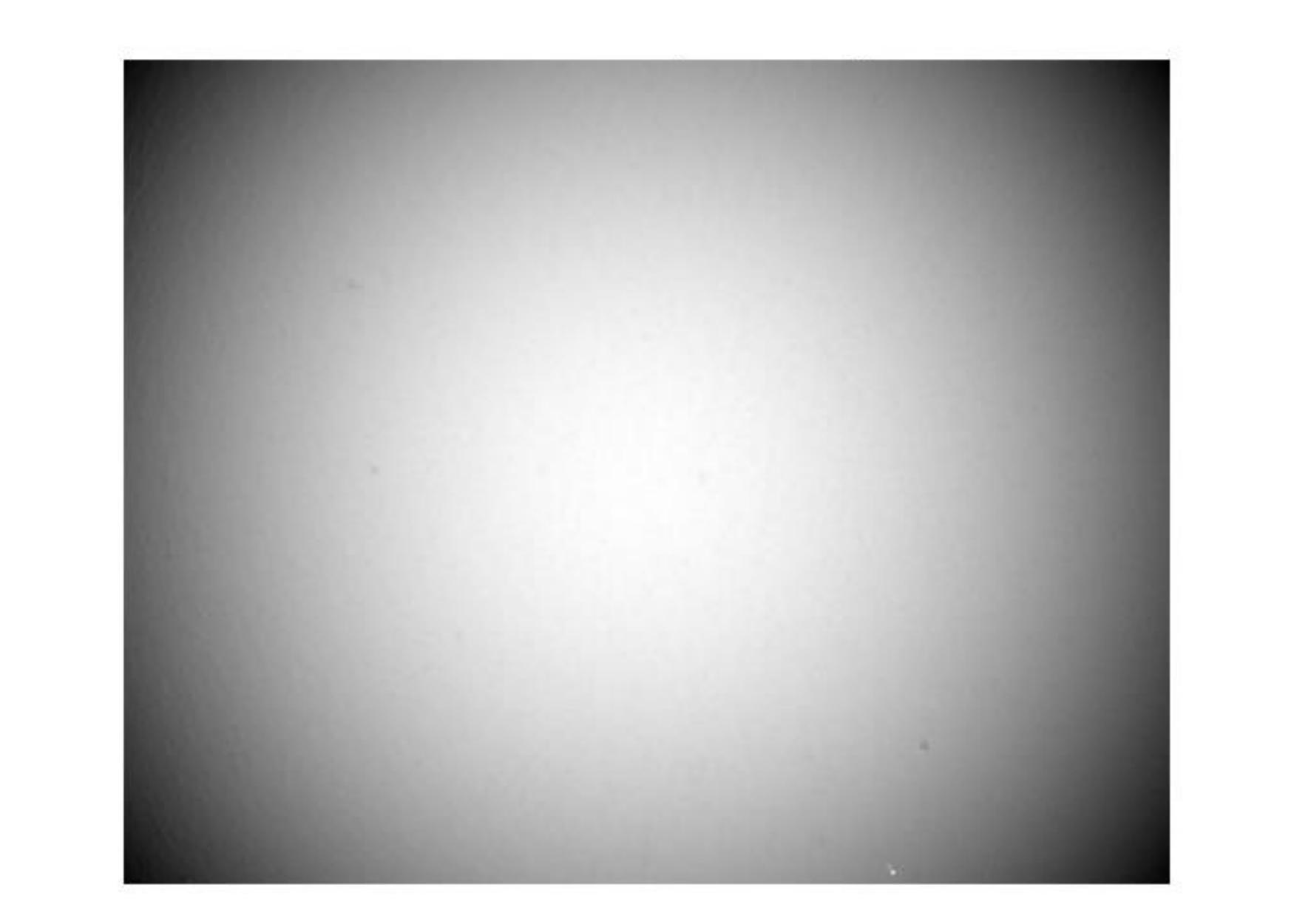}
         \caption{Before pixel height calibration}
     \end{subfigure}
     %\hfill
     \begin{subfigure}[b]{0.49\textwidth}
         \centering
         \includegraphics[width=\textwidth]{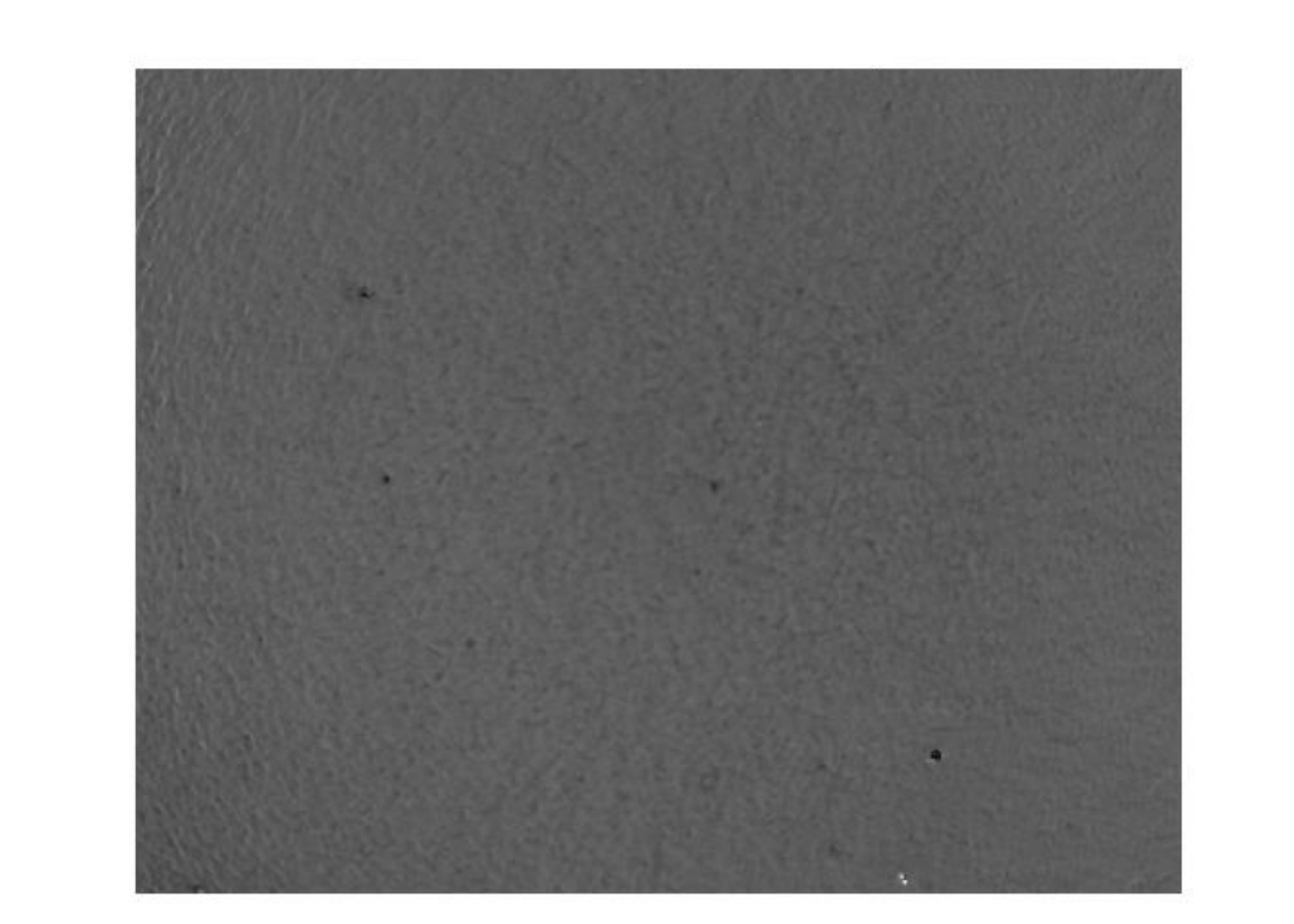}
         \caption{After pixel height calibration}
     \end{subfigure}

     \caption{Visualizations of surface roughness before and after subtracting the surface baseline from the pixel height data.}
     \label{fig:curvature}
\end{figure}

 The original pixel height matrix of $Z$ embodies the inaccurate measurement from the imaging device due to its soft-fixturing process \citep{hulting1995comment}. That is the process that the device takes a few measurements from the actual surface and, based on them, estimates the center location and radius. All following measurements are collected based upon taking the estimation as a reference. \cite{xia2011bayesian} created a figure to illustrate the difference between actual surface and the ``nominal spherical surface'' (that is the surface baseline referenced in this paper) that is a perfectly round circle with the estimated center and radius. However, the inaccurate estimation and the spherical baseline in the pixel height data makes it difficult to discern the surface roughness. One can see Figure \ref{fig:curvature}(a) for the roughness being easily overlooked.

Recall the formula of $Sa$ in Equation \eqref{eq:Sa}.  The position of the surface baseline is characterized by $\bar{z}$.   We undertake the following preprocessing step to adjust the center and radius of the actual surface and recalibrate the height values of the pixels on the surface, so that the subsequent analysis can still follow what was explained in the earlier sections. 

\begin{enumerate}
    \item Convert the matrix $Z$ to a vector $z$. We first read the matrix $Z$ row-wise into a row vector of size $I\times 1$; see Figure \ref{fig:data}(b).  For the problem at hand,  $I=640\times 480 = 307,200$.  Denote the $i$-th element of $z$ by $z_{(i)}, i=1,\cdots,I$. We also re-index $X_w$ and $Y_v$ to become $X_{(i)}$ and $Y_{(i)}$, respectively.  We want to ensure that $z_{(i)}$ is associated with its original coordinates, i.e., those $z_{w,v}$ is associated with. 

    \item Arrange the three newly created row vectors, $X$, $Y$ and $z$ into a matrix as $[X^T, Y^T, z^T]$, where ${X}=[1,\cdots,X_{(I)}]$, $Y=[1,\cdots,Y_{(I)}]$, and $z=[1,\cdots,z_{(I)}]$. Compute the coordinates of the center of the sphere, denoted by $[X_c,Y_c,z_c]$ and the radius of the sphere, denoted by $r$. Such computation can be facilitated by using some software routines, like the \texttt{MATLAB}$^\circledR$ built-in function \texttt{sphereFit}.

    \item Subtract the surface baseline from the vector $z$. Calculate the new pixel height, $z'$, such that \[z'(X_w,Y_v)=\sqrt{(r^2-(X_w-X_c)^2-(Y_v-Y_c)^2)} +z_c.\]
          This new pixel height is calibrated by subtracting the adjusted surface baseline from the original measures. 
\end{enumerate}

 Figure \ref{fig:curvature}(b) displays the surface roughness after the pixel height calibration.  The rough texture of the surface is much more visible than the plot on the left. For the sake of notational simplicity and without ambiguity, we still use $z$ instead of $z'$ to represent the pixel height after the calibration.

 The surface baseline adjustment process could be fairly simple for some geometric shapes, e.g., polyhedral geometries, but may be complicated for the other geometric shapes, e.g., spiral geometries. For the latter case, one may resort to the local regression method to smooth out the surface baseline \citep{cleveland1979robust}. One should note that although the pixel heights are calibrated from the surface baseline, the locations still retain their spatial distances and correlations on their geometrical surface.

\subsection{Study of Bead Polishing Process}
We have conducted two experiments to illustrate the use of the method for detecting, respectively, the lack-of-improvement point during a polishing process and a lapping process. This subsection focuses on the polishing process in the first experiment, whereas Section \ref{sec:lapexp} presents the second experiment, which has only lapping and no polishing action.  Part of the reason that we skip the analysis of lapping in the first experiment is due to insufficient locations sampled on the bead at the lapping stages for our permutation-based method to be applied.

\begin{figure}[h!]
     \centering
     \begin{subfigure}[b]{0.49\textwidth}
         \centering
         \includegraphics[width=\textwidth]{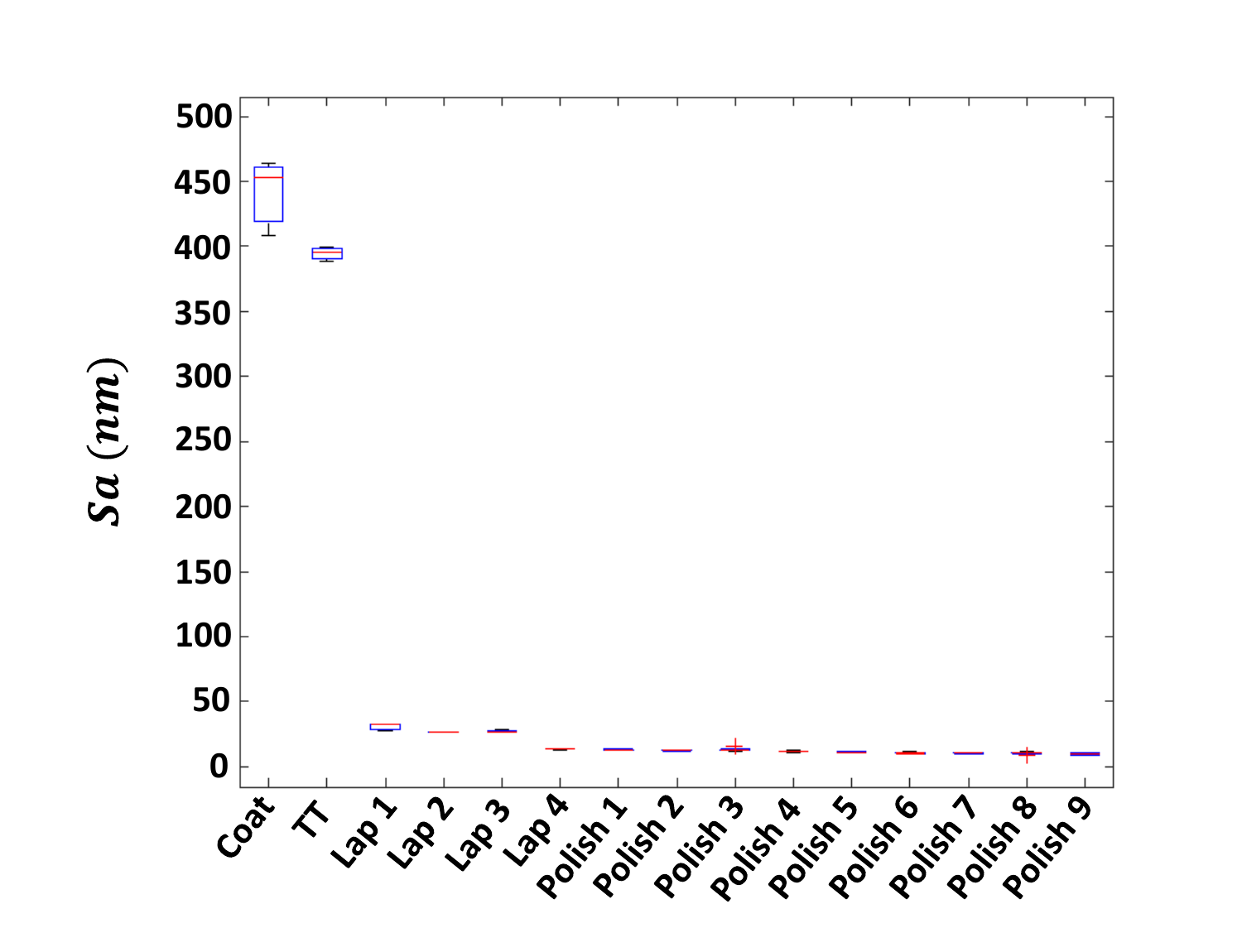}
         \caption{}
     \end{subfigure}
     %\hfill
     \begin{subfigure}[b]{0.49\textwidth}
         \centering
         \includegraphics[width=\textwidth]{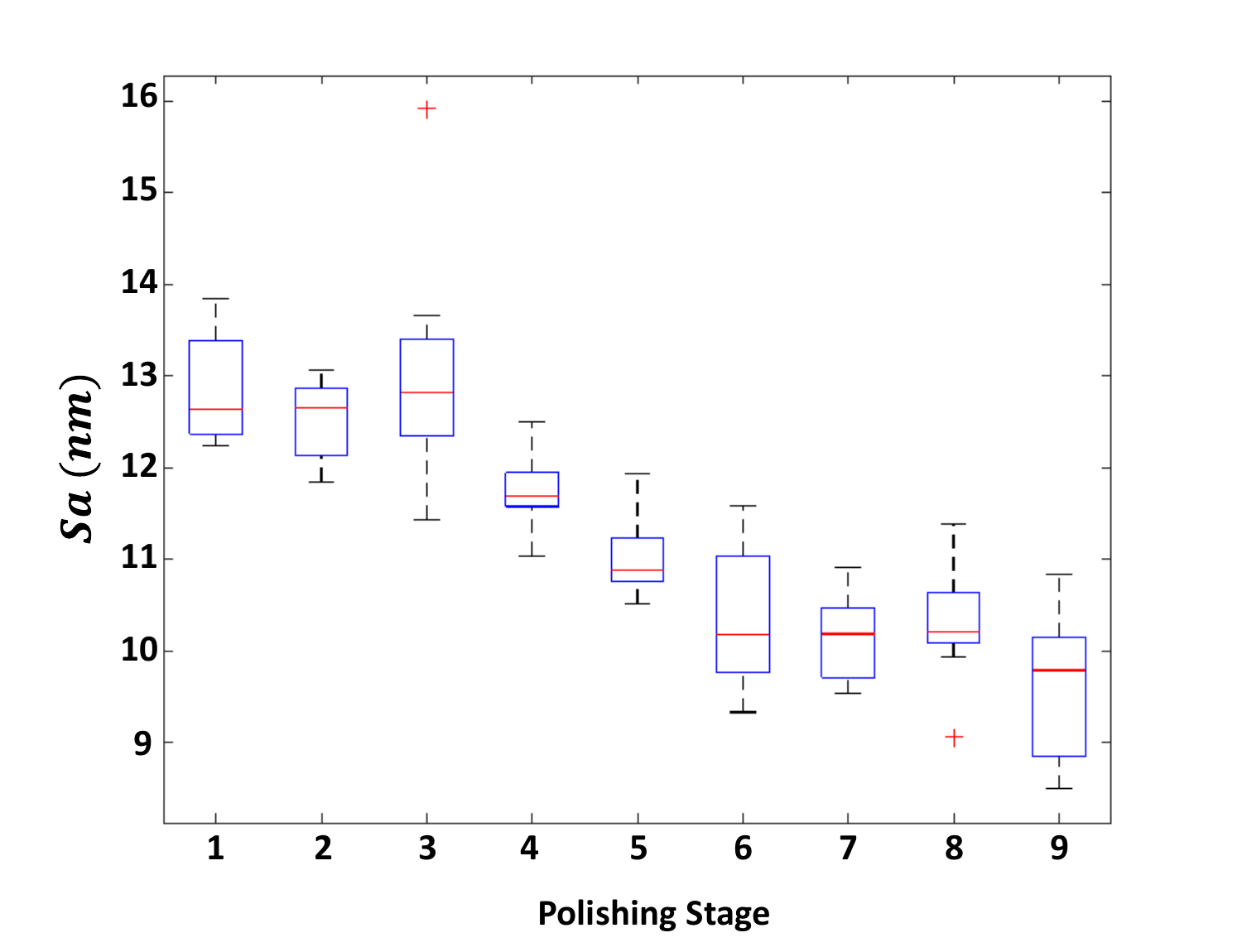}
         \caption{}
     \end{subfigure}

     \caption{Boxplots of $Sa$ values of the ball surface roughness as the surface polishing progresses:(a) shows the $Sa$ values of all processing stages; (b) shows the ones of the nine polishing stages.}
     \label{fig:sa}
\end{figure}

The surface polishing process in the first experiment comprises of a stage of TT, four stages of lapping, and nine stages of polishing. Figure \ref{fig:sa} presents the trend of $\overline{Sa}$ from coating to the last polishing stage. In Figure \ref{fig:sa}(a), the notation along the horizontal axis indicates that the $Sa$ displayed is measured right after that specific processing stage.  For instance, ``Lap $3$'' means the third time when the lapping process was paused and the measurement of the surface roughness was taken. Figure \ref{fig:sa}(b) presents a zoom-in version of the $Sa$ boxplots for the nine polishing stages.

After coating, the $\overline{Sa}$ value of the surface is $453.2$ $nm$. The tip truncating (TT) process reduces the $\overline{Sa}$ value to $395.7$ $nm$. Afterward, the four stages of lapping process bring the value significantly down to an average of $24.8$ $nm$, suggesting that the lapping process makes significant progress in smoothing out the initial rough surface. Up to this point, the use of $\overline{Sa}$ is adequate and there is hardly any disagreement on decision making or action taking.

From the first stage of polishing process and onward, however, the reduction in $\overline{Sa}$ is not that significant. The zoom-in view on Figure \ref{fig:sa}(b) shows some degree to further reduction, from roughly $12.5$ $nm$ to a little bit over $10$ $nm$.  But the cost to accomplish that reduction is huge.  From Polish $1$ to Polish $9$, the time spend is a total of $240$ hours, or $10$ days.  Under such long time of polishing, the risk has increased considerably that the bead, which is hallow inside, could crack, or other damages may happen to it. It would be preferred if such overpolishing can be avoided by triggering an earlier stopping point.

\subsubsection{Hypothesis Tests Results versus $Sa$}
Table \ref{tab:phyresults} presents the results using the hypothesis testing-based decision process.  It includes the p-values associated with the three hypothesis tests and the decisions that the $p$-value suggests. Recall that the detection rule is that as long as any of the null hypotheses is rejected, the surface quality is considered being improved beyond random fluctuation.  If none of the null hypotheses is rejected, we deem that no significant surface quality improvement has been detected.

To register a decision, a threshold for $p$-value is needed. However, the $p$-value thresholds chosen in practice, for instance, the customary $0.05$ cut-off, bears certain degree of arbitrariness.  A closer look at the $p$-values in Table \ref{tab:phyresults} offers a clue in selection.  It is apparent that the $p$-values therein belong to three groups, $\{0.02, 0.04, 0.05, 0.06\}$, $\{0.10, 0.12\}$, and $\{ > 0.22\}$.  Recall that we have three hypothesis tests in each decision.  By invoking Bonferroni adjustment, the type-I error for the overall decision is three times of that of individual hypothesis tests. Should we deem a $p$-value of $0.10$ or $0.12$ significant at the individual test level, it implies the type-I error for the overall decision is at the level of $0.30$ or greater.  This seems outside the typical choice of $p$-value cut-off.  For this reason, we deem that only the $p$-values in the first group imply certain degrees of significance. The corresponding type-I error at the overall decision level is therefore $\{0.06, 0.12, 0.15, 0.16\}$, respectively.  We further treat that a $p$-value smaller than $0.033$ at the individual test level, or smaller than $0.1$ at the overall decision level, as significant, and a $p$-value between $0.033$ and $0.067$ at the individual test level, or between $0.1$ and $0.2$ at the overall decision level, as marginally significant. In the other words, the significance levels used here are $90\%$ and $80\%$, so that a decision is significant when the associated $p$-value is greater than $90\%$, insignificant when lower than $80\%$, and marginally significant for those in between.

\begin{table}[h!]
\begin{center}
\caption{Test results on the nine polishing stages using the hypothesis testing method. $P\#$ in the \textit{Stages} column indicates a specific polishing stage.}
\label{tab:phyresults}
\footnotesize
\begin{tabular}{  c | c c | c c | c c | c }
\hline \hline
		 & \multicolumn{2}{c|}{Mean test for} & \multicolumn{2}{c|}{Mean test for} & \multicolumn{2}{c|}{Variance test} & \\
        & \multicolumn{2}{c|}{upper tail} &
        \multicolumn{2}{c|}{lower tail} &  \multicolumn{2}{c|}{} &  Surface quality \\
\cline{2-7}
Stages & p-value & Outcome & p-value & Outcome & p-value & Outcome &  improvement detection\\
\hline
\hline
P$1$ vs. P$2$& $0.58$ & Not lowered & $0.22$ & Not raised & $0.28$ & Not reduced & No improvement detected\\
\hline
P$2$ vs. P$3$ & $0.89$ & Not lowered & $0.94$ & Not raised & $0.82$  & Not reduced & No improvement detected\\
\hline
P$3$ vs. P$4$ & $0.65$ & Not lowered & $0.98$ & Not raised & $0.02$ & Reduced & Improvement detected\\
\hline
P$4$ vs. P$5$ & $0.10$ & Not lowered & $0.27$ & Not raised & $0.32$ & Not reduced & No improvement detected\\
\hline
P$5$ vs. P$6$ & $0.06$ &
\begin{tabular}{@{}c@{}}Lowered \\ (marginal)\end{tabular}  & $0.28$ & Not raised & $0.92$ & Not reduced & \begin{tabular}{@{}c@{}}Improvement detected\\(marginal)\end{tabular} \\
\hline
P$6$ vs. P$7$ & $0.12$ & Not lowered & $0.74$ & Not raised & $0.05$ & \begin{tabular}{@{}c@{}}Reduced\\(marginal)\end{tabular} & \begin{tabular}{@{}c@{}}Improvement detected\\(marginal)\end{tabular} \\
\hline
P$7$ vs. P$8$ & $0.96$ & Not lowered & $0.91$ & Not raised & $0.96$ & Not reduced & No improvement detected \\
\hline
P$8$ vs. P$9$ & $0.04$ & \begin{tabular}{@{}c@{}}Lowered \\(marginal)\end{tabular}& $0.81$ & Not raised & $0.46$ & Not reduced & \begin{tabular}{@{}c@{}}Improvement detected\\(marginal)\end{tabular} \\
\hline
\hline
\end{tabular}
\end{center}
\end{table}

The hypothesis testing procedure detects a reduced variance from Stage P$3$ to Stage P$4$, and to a lesser degree of significance, a lowered upper tail from Stage P$5$ to Stage P$6$, a reduced variance from Stage P$6$ to Stage P$7$ and a lowered upper tail from Stage P$8$ to Stage P$9$. These detection outcomes show a good consistency with observing the $Sa$ boxplots in Figure \ref{fig:sa}(b). The detection of reduced variance is more consistent between the hypothesis tests and the $Sa$ boxplots. The hypothesis tests signal two instances of reduced variance, from Stage P$3$ to Stage P$4$ and then from Stage P$6$ to Stage P$7$, both of which are visibly so in the $Sa$ boxplots. Even though these outcomes are consistent, the merit of using the hypothesis testing method is that the new method provides a more detailed information informing users about where a change is detected and to which degree it is significant.

On the other hand, not all the detection outcomes using the hypothesis testing method is the same as using $Sa$.  Using the $Sa$ values appears to be more readily in signaling an improvement in the surface quality. Consider the decreasing trend of $\overline{Sa}$ from Stage P$3$ to Stage P$6$ in Figure \ref{fig:sa}(b).  By contrast, the hypothesis tests do not detect significant improvement on the surfaces after these polishing stages, per criteria defined in Equations (\ref{eq:dis_h1})--(\ref{eq:dis_h3})), except that some marginal improvement at the upper tail (i.e., peak removals) from P$5$ to P$6$.

\begin{figure}[p]
     \centering
     %\vspace{1mm}
     \begin{subfigure}[b]{\textwidth}
         \centering
         \includegraphics[width=\textwidth]{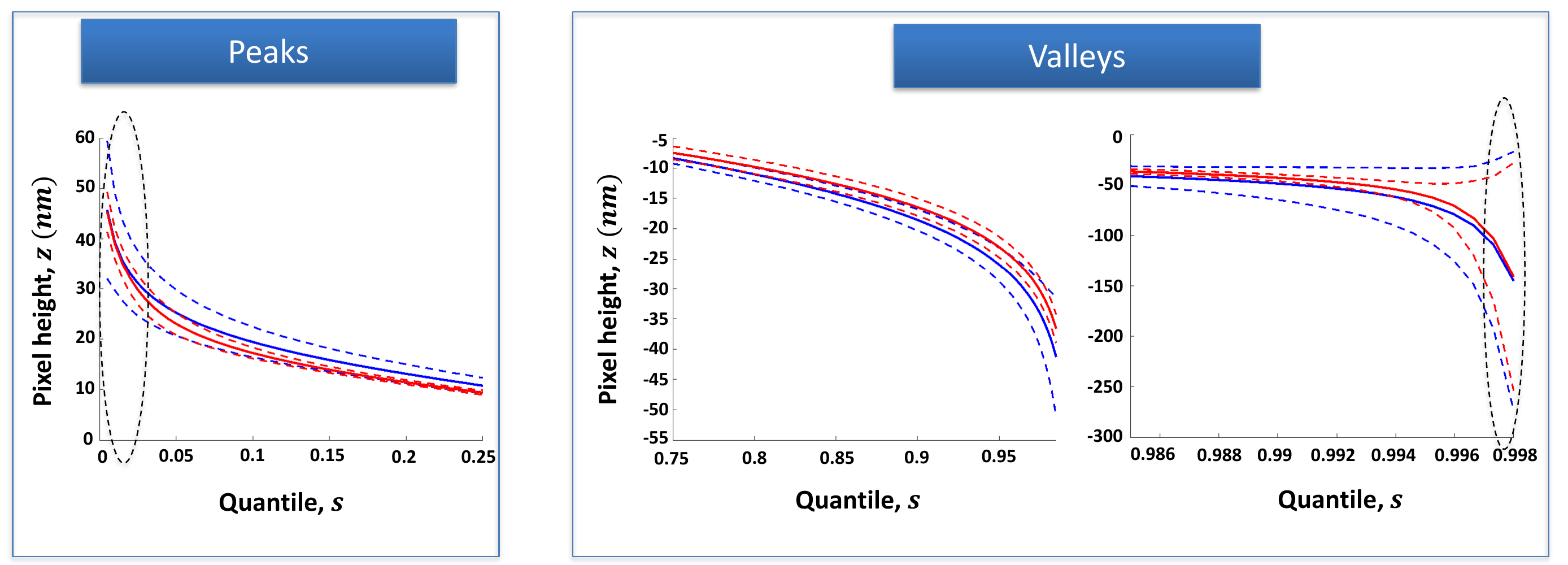}
         \caption{}
     \end{subfigure}
     %\vspace{1mm}
     \begin{subfigure}[b]{\textwidth}
         \centering
         \includegraphics[width=\textwidth]{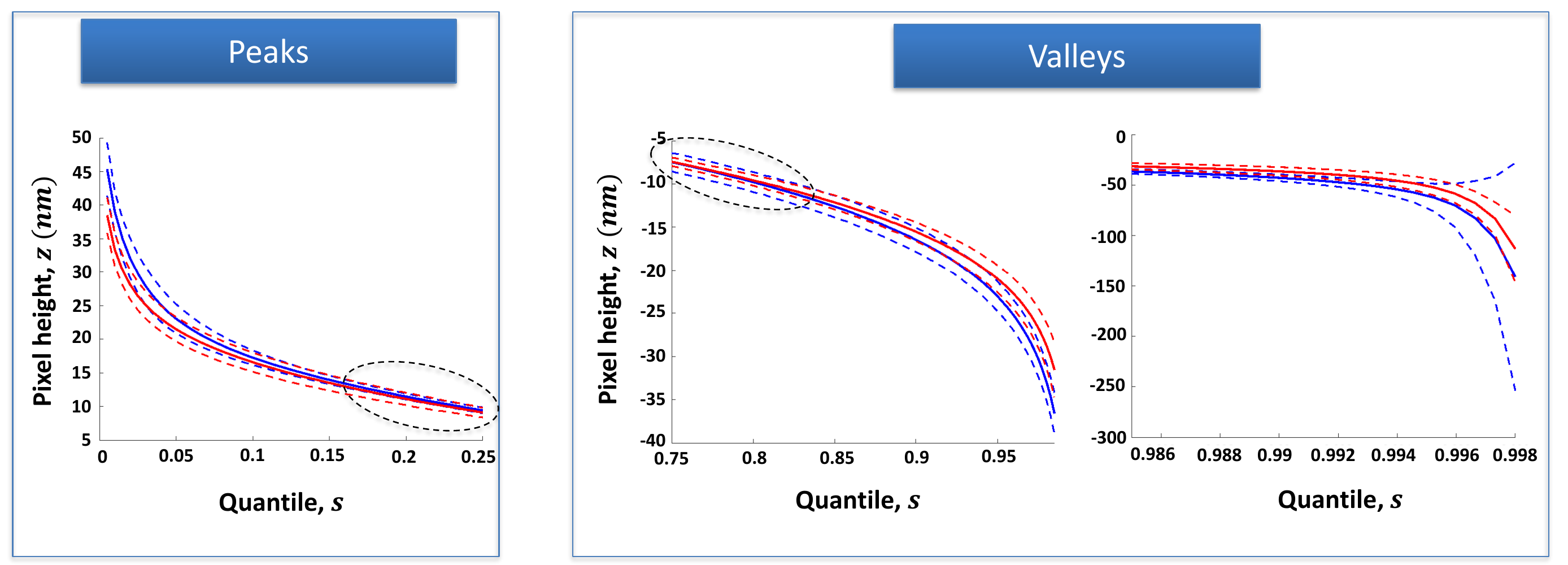}
         \caption{}
     \end{subfigure}
     %\vspace{1mm}
     \begin{subfigure}[b]{\textwidth}
         \centering
         \includegraphics[width=\textwidth]{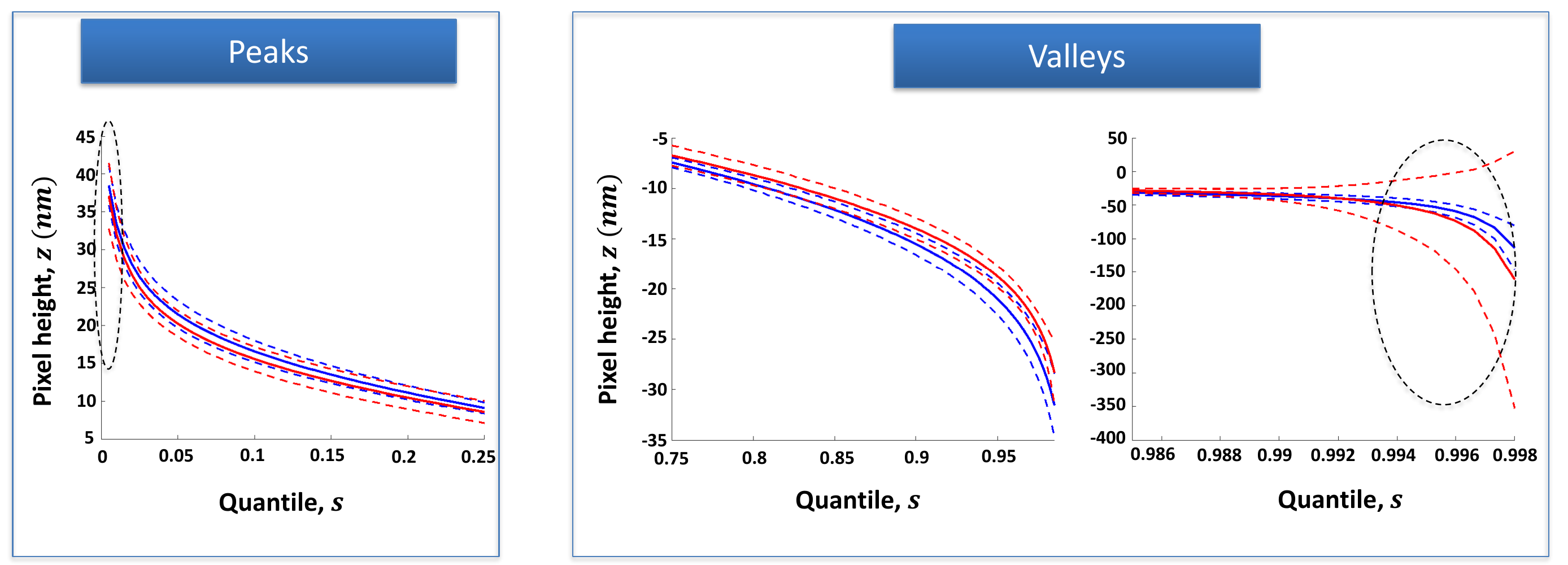}
         \caption{}
     \end{subfigure}
     \vspace{2mm}
    \begin{subfigure}[b]{0.4\textwidth}
         \centering
         \includegraphics[width=\textwidth]{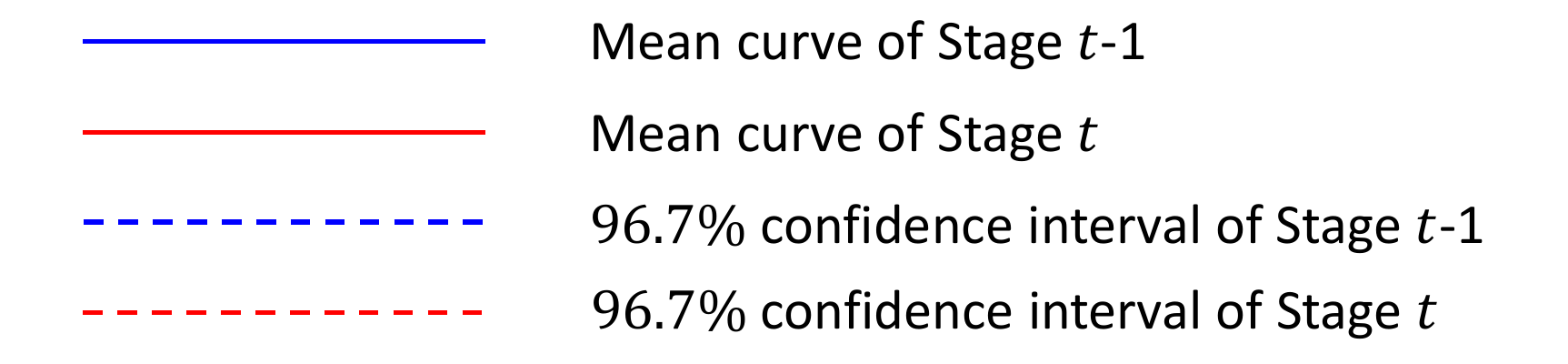}
         %\caption{}
     \end{subfigure}

     \caption{Mean curves (solid curves) and $96.7\%$ confidence bands (dashed curves) of BACs at two consecutive stages: (a) Polish $3$ vs. Polish $4$; (b) Polish $4$ vs. Polish $5$ and (c) Polish $5$ vs. Polish $6$. }
     \label{fig:BACs4comp}
\end{figure}

In Figure \ref{fig:BACs4comp}, we plot the average BACs and the associated $96.7\%$ confidence bands ($96.7\%$ corresponds to the $0.33$ $p$-value cut-off) on two consecutive stages to visualize the two-stage surface roughness changes and discern if the outcome difference between the hypothesis tests and $Sa$ makes sense. Since the median $Sa$ shows a decreasing trend, we hope to see that (1) the mean curve of Stage $t$ (solid red curve) representing the peaks (Quantile $0$ to $0.25$) is lowered compared to that of Stage $t-1$ (solid blue curve) and/or (2) the mean curve of Stage $t$, representing the valleys (Quantile $0.75$ to $1$) is raised. The reality is that some of the tail portions of the BACs does not show significant differences.  Even if there are some differences, those appear to be well within the confidence bands.  We mark certain areas in Figure \ref{fig:BACs4comp} with dashed circles to highlight. It is evident to us that the median $Sa$ fails to capture such regional differences but these regional differences are crucial in reaching a sensible decision for these ultra-precise manufacturing products.

We also observe that once in a while, the BACs at a later stage of polishing could deteriorate as compared with the preceding stage. Consider the example of the lower tail portion from Stage P$5$ to Stage P$6$ (Figure \ref{fig:BACs4comp} (c)), representing their valleys. It turns out that after additional polishing the valleys at Stage P$6$ are deeper than those at Stage P$5$, mostly likely due to extra scratches introduced on the surface during polishing, which is a clear indication of over polishing. Please note that in our plot, we only plot the Quantile $0.75$ to $0.998$ while leaving out a tiny portion close to $1$, because that extremely deep valley pixels are sometimes of much great depth and their presence compresses the scale on the plot, making the visualization difficult.

If using the hypothesis testing outcomes in Table \ref{tab:phyresults}, the polishing process should stop at Stage P$2$, or if not so quickly, at the latest at Stage P$5$.  Doing either will save significant polishing time, reduce materials removal, and lower the risk of over polishing and bead damaging in the long operation stretch.

\subsection{Study of Bead Lapping Process}\label{sec:lapexp}
\begin{figure}[t]
     \centering
     \begin{subfigure}[b]{0.45\textwidth}
         \centering
         \includegraphics[width=\textwidth]{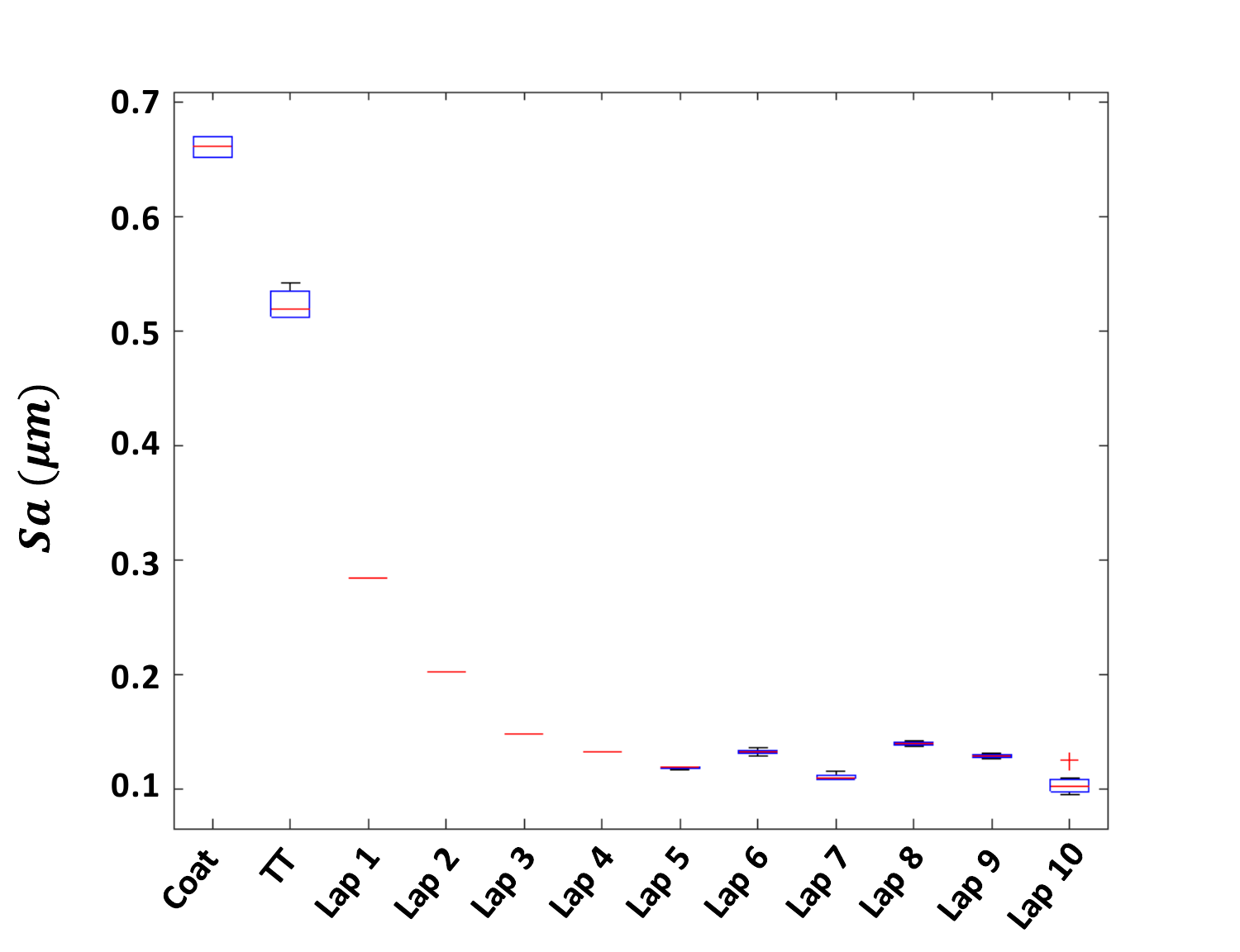}
         \caption{}
        %  \label{}
     \end{subfigure}
     \hfill
     %\hspace{15mm}
     \begin{subfigure}[b]{0.46\textwidth}
         \centering
         \includegraphics[width=\textwidth]{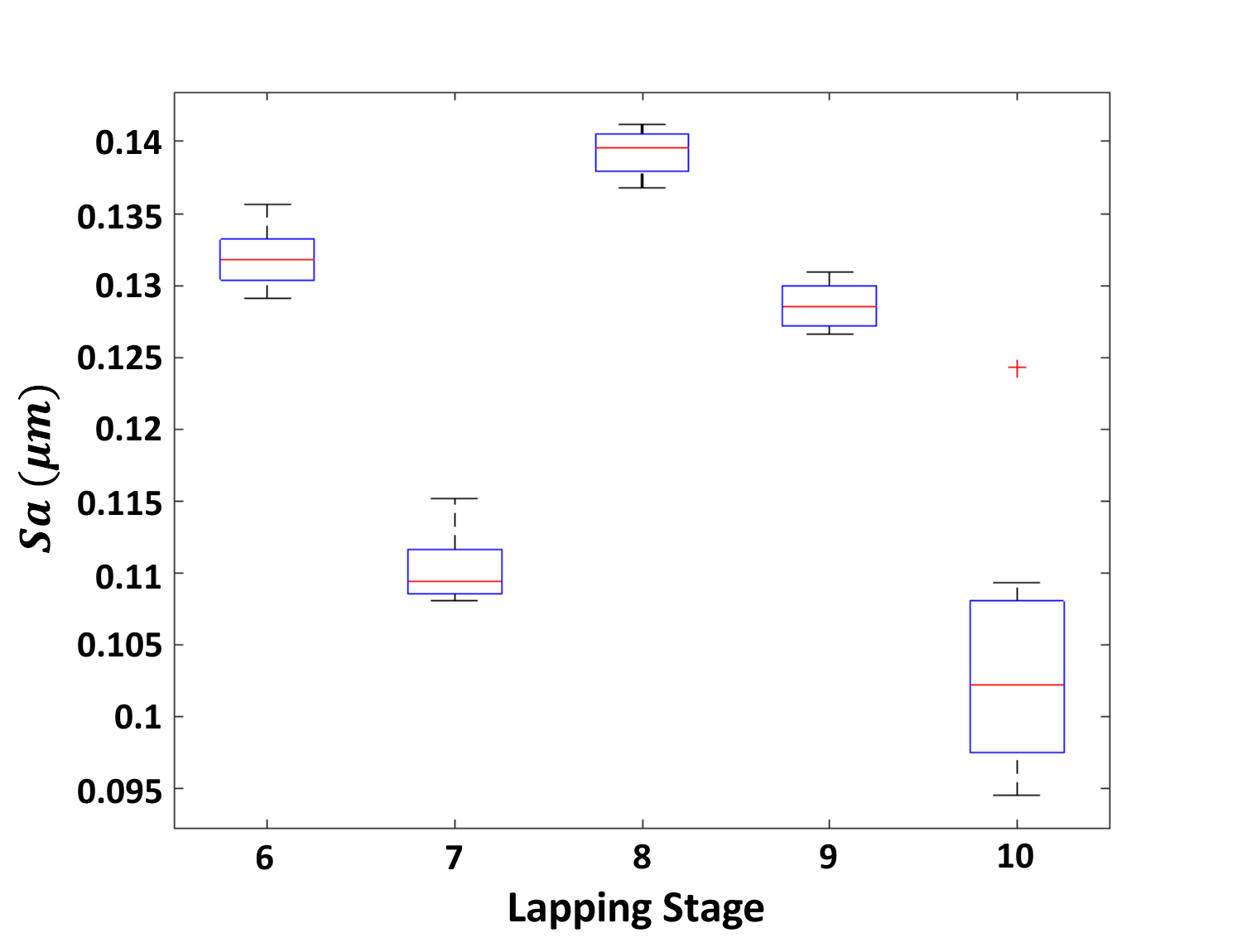}
         \caption{}
        %  \label{}
     \end{subfigure}
     \caption{$Sa$ boxplots for the second physical experiment. (a) $Sa$ boxplots for all stages; (b) $Sa$ boxplots for Stage L$6$ and onward. }
     \label{fig:secondexp}
\end{figure}

The second experiment comprises a stage of coating, a stage of TT and ten stages of lapping.  Figure \ref{fig:secondexp} presents the boxplots of $Sa$ for the whole process, where Figure \ref{fig:secondexp}(b) is the zoom-in view of Stage L$6$ and onwards.  The lapping process, after L$5$, is able to bring the surface roughness to the level of $12$ $nm$.  This time, the pattern of fluctuation in $Sa$ is much more obvious, making the decision harder using the traditional decision tools.  From Stage L$6$ and onward, the time spent, a total of $80$ hours, and the material removed, $7.72$ $mg$, are again significant.

\begin{figure}[t]
     \centering
      \begin{subfigure}[b]{0.8\textwidth}
         \centering
         \includegraphics[width=\textwidth]{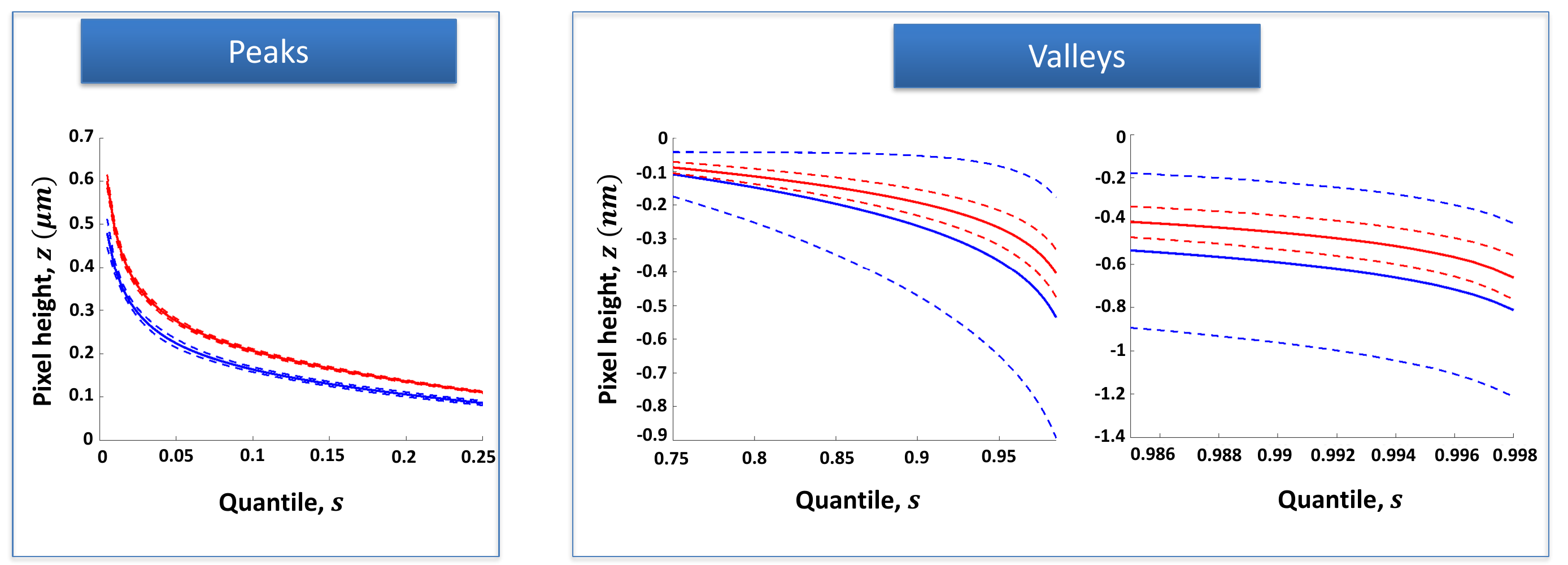}
         \caption{}
        %  \label{}
     \end{subfigure}
     \hfill
     %\hspace{15mm}
     \begin{subfigure}[b]{\textwidth}
         \centering
         \includegraphics[width=0.4\textwidth]{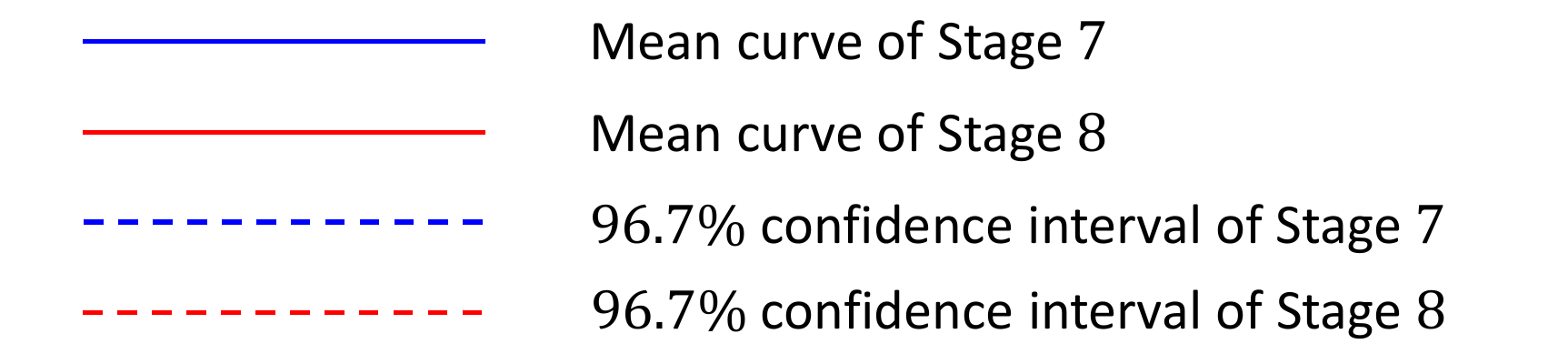}
         %\caption{}
        %  \label{}
     \end{subfigure}
     \caption{Mean curves (solid curves) and $96.7\%$ confidence intervals (dashed curves) of the BACs of Lap $7$ vs. Lap $8$.}
     \label{fig:l7vsl8}
\end{figure}

The hypothesis tests results shown in Table \ref{tab:secondresults} detect the variance reduction from Stage L$7$ to Stage L$8$, but do not confirm the mean reduction trend as observed in the $Sa$ plot for Stage L$8$ to L$10$, except for a marginal peak flattening from L$8$ to L$9$. Inspecting the $Sa$ boxplots, the variance reduction from Stage L$7$ to Stage L$8$ is not so obvious. We therefore present the detailed BACs and the associated confidence bands in Figure \ref{fig:l7vsl8} for Stage L$7$ (blue curves) versus L$8$ (red curves). On these plots, it is clearly showing that the band of L$8$ is narrower than that of L$7$.

\begin{table}[h]
\begin{center}
\caption{Test results of L$6$ and onward. $L\#$ in the \textit{Stages} column indicates a specific lapping stage in the process.}
\label{tab:secondresults}
\footnotesize
\begin{tabular}{  c | c c | c c | c c | c }
\hline \hline
		 & \multicolumn{2}{c|}{Mean test for} & \multicolumn{2}{c|}{Mean test for} & \multicolumn{2}{c|}{Variance test} & \\
        & \multicolumn{2}{c|}{upper tail} &
        \multicolumn{2}{c|}{lower tail} &  \multicolumn{2}{c|}{} &  Surface quality \\
\cline{2-7}
Stages & p-value & Outcome & p-value & Outcome & p-value & Outcome &  improvement detection\\
\hline
\hline
L$6$ vs. L$7$ & $0.33$ & Not lowered & $0.14$ & Not raised & $0.89$ & Not reduced & No improvement detected \\
\hline
L$7$ vs. L$8$ & $0.99$ & Not lowered & $1$ & Not raised & $2\times 10^{-5}$ & Reduced & Improvement detected \\
\hline
L$8$ vs. L$9$ & $0.06$ & \begin{tabular}{@{}c@{}}Lowered \\(marginal)\end{tabular} & $0.94$ & Not raised & $0.83$ & Not reduced & \begin{tabular}{@{}c@{}}Improvement detected\\(marginal)\end{tabular} \\
\hline
L$9$ vs. L$10$ & $0.41$ & Not lowered & $0.99$ & Not raised & $0.83$ & Not reduced & No improvement detected \\
\hline
\hline
\end{tabular}
\end{center}
\end{table}

\begin{figure}[p]
     \centering
     \begin{subfigure}[b]{\textwidth}
         \centering
         \includegraphics[width=\textwidth]{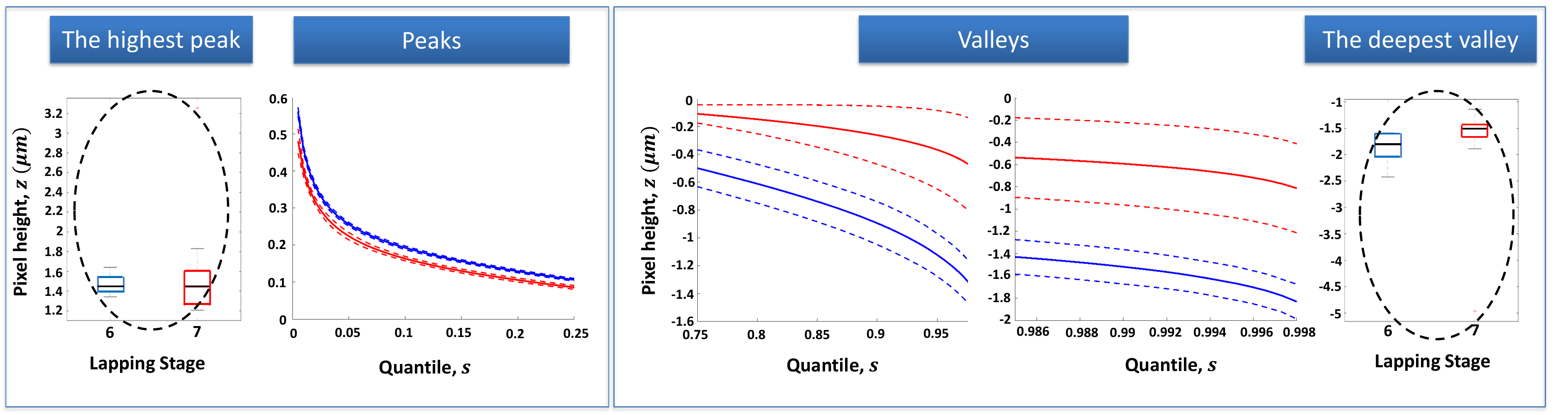}
         \caption{}
        %  \label{}
     \end{subfigure}
     \hfill
     %\hspace{15mm}
     \begin{subfigure}[b]{\textwidth}
         \centering
         \includegraphics[width=\textwidth]{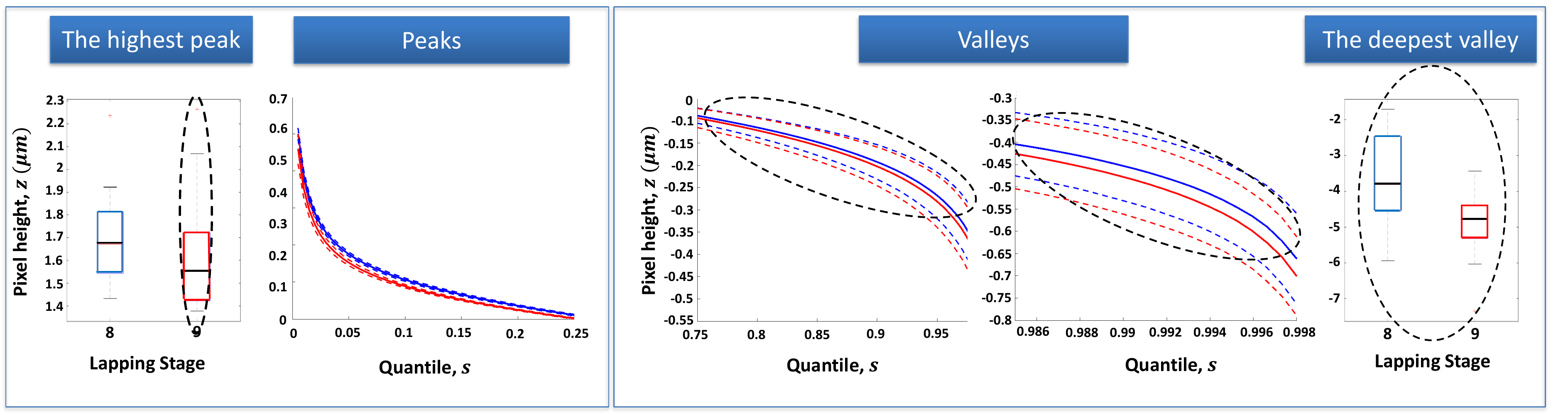}
         \caption{}
        %  \label{}
     \end{subfigure}
     \begin{subfigure}[b]{\textwidth}
         \centering
         \includegraphics[width=\textwidth]{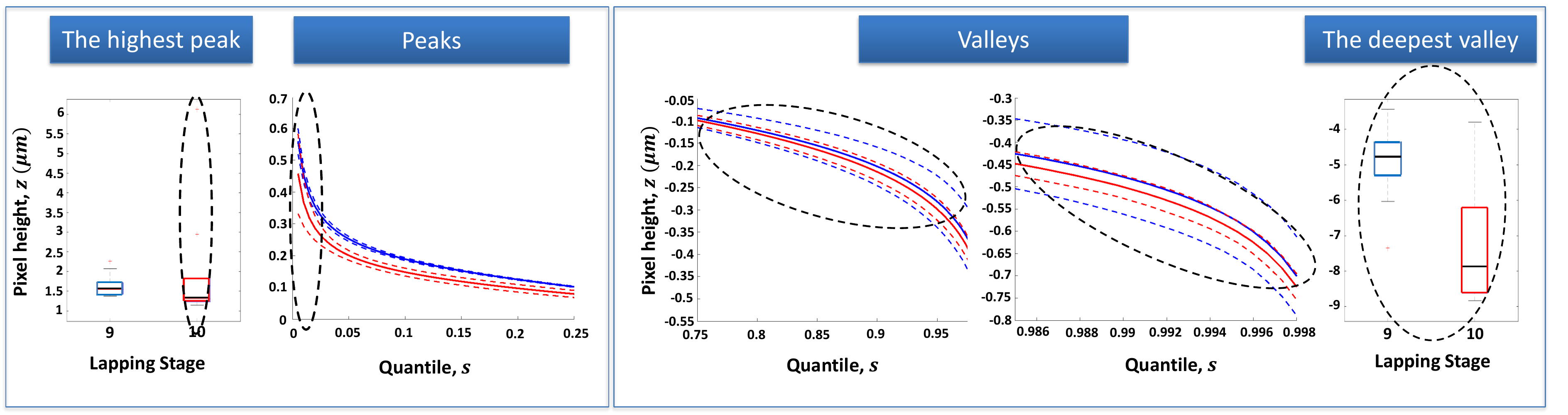}
         \caption{}
        %  \label{}
     \end{subfigure}
     \begin{subfigure}[b]{\textwidth}
         \centering
         \includegraphics[width=0.4\textwidth]{Figure7f.pdf}
         %\caption{}
        %  \label{}
     \end{subfigure}
     \caption{Mean curves (solid curves) and $6\sigma$ confidence bands (dashed curves) of the bearing area curves for two consecutive stages for Stage L$6$ and onward: (a) Lap $6$ vs. Lap $7$; (b) Lap $8$ vs. Lap $9$; (c) Lap $9$ vs. Lap $10$.}
     \label{fig:secondBACs}
\end{figure}

Figure \ref{fig:secondBACs} presents BACs for Stage L$6$ and onward and compare those on two consecutive stages. The mean curve comparison show that between Stages L$6$ and L$7$, the reason for a lack of detection by the hypothesis testing method is due to that either the highest peak or the deepest valley is not improved.  Some may argue that those are outliers and should not be considered, while others may argue that the extreme of the peaks and valleys does reflect the surface roughness, as they reveal the bumps and scratches produced during the action of lapping.  We want to note that our decision process can be easily tailored to suit different needs in the specific context of applications.  In case that the sensitivity to the peaks and valleys is appreciated, then the above results show that the hypothesis tests do have the desired sensitivity in detection.  If the highest peaks or the deepest valleys should be excluded from decision making, then one just needs to adjust the definition of $D_u$ and $D_l$ in Equations (\ref{eq:dis_h1})--(\ref{eq:dis_h3})) to accommodate such changes.  The rest of the testing procedure stays more or less the same.

\subsection{Detection Sensitivity Analysis}
In this section, we perform sensitivity studies on the number of sampling locations in Section \ref{sec:sennbofloc} and on the value of $\tau$ in Section \ref{sec:sentau}.  Recall that $\tau$ specifies the tail portion used in Equations (\ref{eq:dis_h1})--(\ref{eq:dis_h3}). 

\subsubsection{Sensitivity on Number of Sampling Locations}\label{sec:sennbofloc}
The experiments leading to the hypothesis test results in Tables \ref{tab:phyresults} and \ref{tab:secondresults} are based on nine sampling locations, randomly sampled on the bead's surface.  We do not recommend using fewer than nine sampling locations as our decision procedure is nonparametric in nature and based on permutation. When there are too few curves, i.e., too few sampling locations, a permutation will easily repeat the same combination and thus becomes less effective.  On the other hand, doing in-process measurements is costly, especially on a small product like the beads in our process.  We wonder whether nine sampling locations are sufficient in reaching a robust conclusion.

As lapping and polishing are disruptive operations, once operated, the surface cannot be restored to its original state to take more measurements.  To support a sensitivity study, we therefore conduct a new experiment with more sampling locations taken during the process. This is a lapping process with five operating stages, labeled as L1 through L5.  Please note that because this is a brand-new experiment, these lapping stages are not directly comparable to the lapping or polishing stages in the earlier sections. For each operating stage, we divide the bead surface into three shells and randomly sample at three, four, and five locations on each shell, respectively. In total, the number of the sampling locations on the bead's surface are nine, twelve, and fifteen locations, respectively. The results using different sampling locations are shown in Table \ref{tab:senanalnbofloc}.

\begin{table}[h!]
\begin{center}
\caption{A sensitivity analysis of the number of sampling locations. }
\label{tab:senanalnbofloc}
\footnotesize
\begin{tabular}{  c | c | c | c | c | c | c | c | c  }
\hline \hline
  &  & \multicolumn{2}{c|}{Mean test for} & \multicolumn{2}{c|}{Mean test for} & \multicolumn{2}{c|}{} & \\
  %\cline{3-8}
		 &  &\multicolumn{2}{c|}{upper tail} & \multicolumn{2}{c|}{lower tail} & \multicolumn{2}{c|}{Variance test} & Decision \\
\cline{3-8}
 \# of locations &Stages & p-value & outcome  & p-value  & outcome &p-value &outcome & suggested \\
\hline
\multirow{4}{*}{$9$} & L$1$ vs. L$2$ & $0.8429$  & Not lowered & $0.9249$  & Not raised &$0.0035$ & Reduced & Continue \\
  & L$2$ vs. L$3$ & $0.3794$ & Not lowered & $0.0447$ & Not raised & $0.0185$ & Reduced & Continue \\
  & L$3$ vs. L$4$ & $0.0004$ & Lowered & $2\times10^{-5}$ & Raised & $1$ & Not reduced & Continue \\
   & L$4$ vs. L$5$ & $1$ & Not lowered & $1$ & Not raised & $2\times10^{-5}$ & Reduced & Continue \\
 \hline
\multirow{4}{*}{$12$} & L$1$ vs. L$2$ & $0.5783$ & Not lowered & $0.9254$ & Not raised & $0.0001$ & Reduced & Continue \\
& L$2$ vs. L$3$ & $0.3461$ & Not lowered & $0.1291$ & Not raised & $0.1576$ & Not reduced & Stop \\
& L$3$ vs. L$4$ & $0.0135$ & Lowered & $2\times10^{-5}$ & Raised & $1$ & Not reduced & Continue \\
& L$4$ vs. L$5$ & $1$ & Not lowered & $1$ & Not raised & $2\times10^{-5}$ & Reduced & Continue \\
\hline
\multirow{4}{*}{$15$} & L$1$ vs. L$2$ & $0.5333$ & Not lowered & $0.9414$ & Not raised & $6\times10^{-5}$ & Reduced & Continue \\
& L$2$ vs. L$3$ & $0.3473$ & Not lowered & $0.3497$ & Not raised & $0.3659$ & Not reduced & Stop \\
& L$3$ vs. L$4$ & $0.0055$ & Lowered & $2\times10^{-5}$ & Raised & $1$ & Not reduced & Continue \\
& L$4$ vs. L$5$ & $1$ & Not lowered & $1$ & Not raised & $2\times10^{-5}$ & Reduced & Continue \\
\hline
\hline
\end{tabular}
\end{center}
\end{table}

\begin{figure}[h]
     \centering
     \begin{subfigure}[b]{0.32\textwidth}
         \centering
         \includegraphics[width=\textwidth]{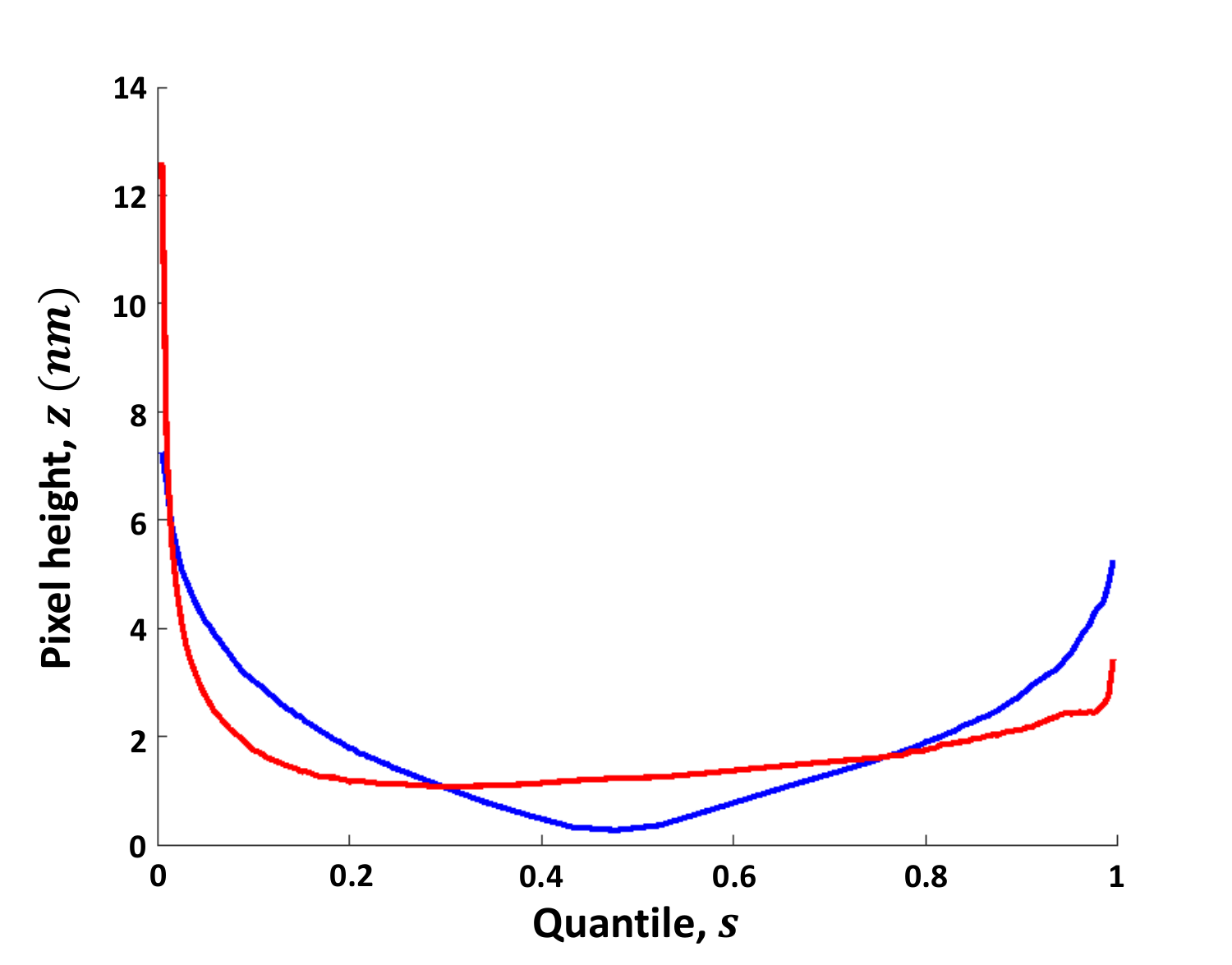}
         \caption{}
        %  \label{}
     \end{subfigure}
     \begin{subfigure}[b]{0.32\textwidth}
         \centering
         \includegraphics[width=\textwidth]{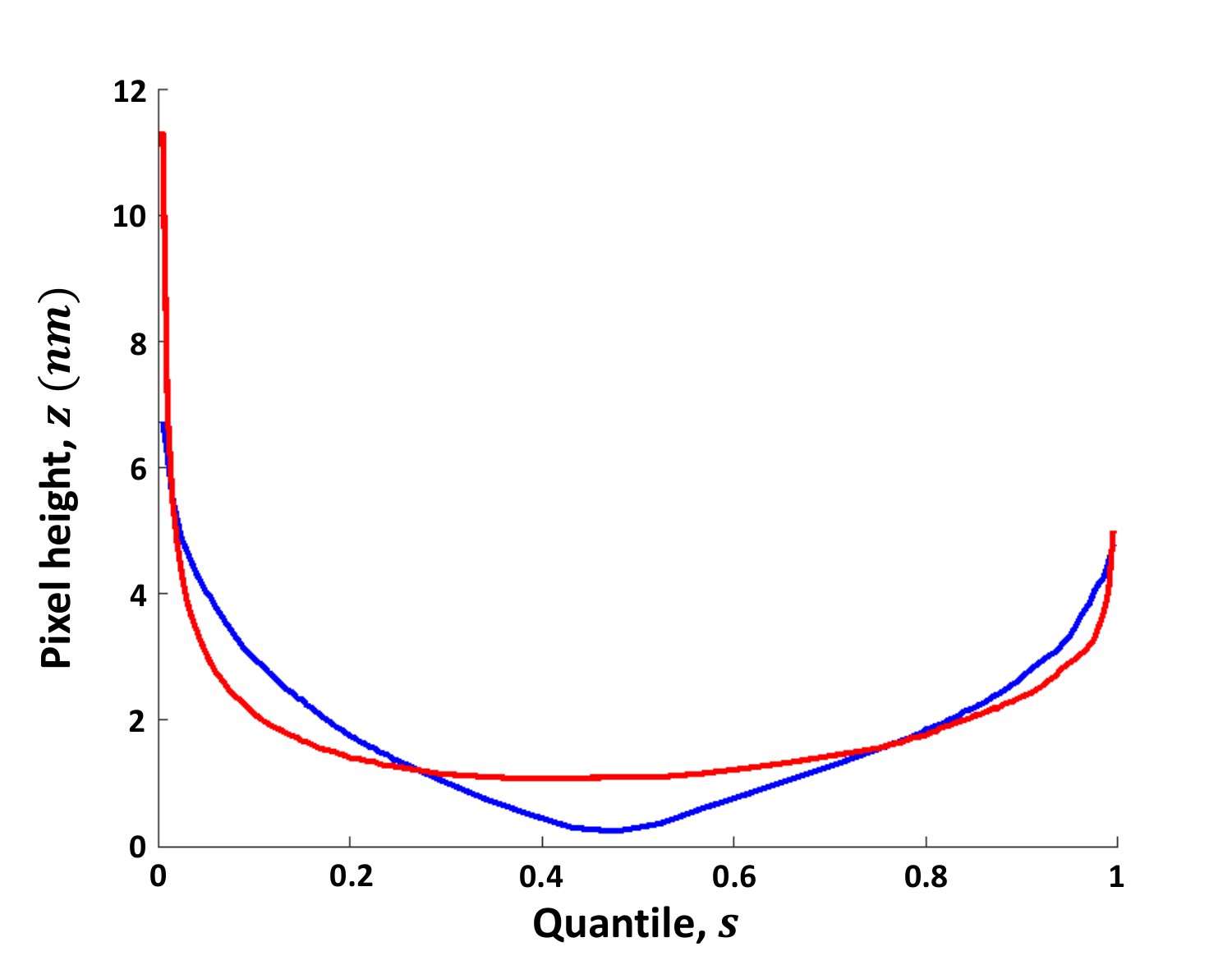}
         \caption{}
        %  \label{}
     \end{subfigure}
     \begin{subfigure}[b]{0.32\textwidth}
         \centering
         \includegraphics[width=\textwidth]{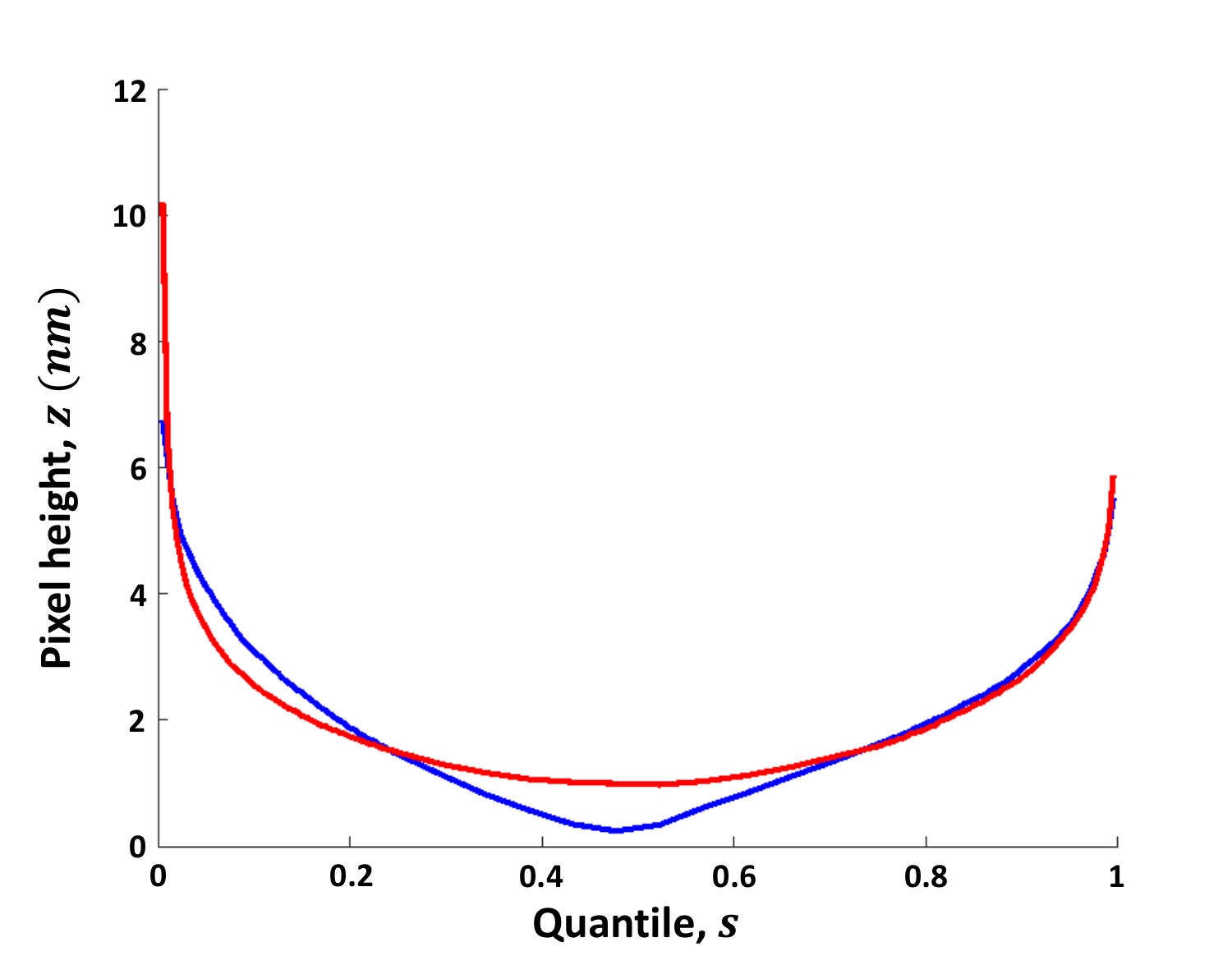}
         \caption{}
        %  \label{}
     \end{subfigure}
     \begin{subfigure}[b]{0.35\textwidth}
         \centering
         \includegraphics[width=\textwidth]{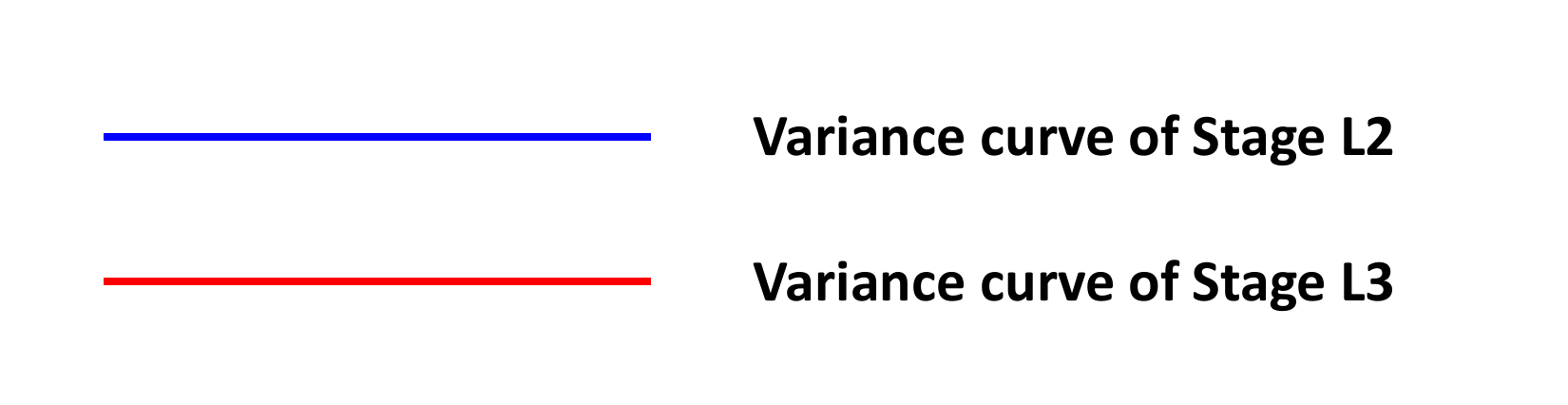}
         %\caption{}
        %  \label{}
     \end{subfigure}
     \caption{Variance curves of BACs of Stage L$2$ (blue curve) vs. L$3$ (red curve) for $9$, $12$ and $15$ sampling locations. (a) Number of locations = $9$; (b) Number of locations = $12$; (c) Number of locations = $15$. }
     \label{fig:varcurves}
\end{figure}

We make two observations.  The first observation is that the overall conclusion is reasonably consistent and stable when using nine or more sampling locations---other than the variance test of Stage L$2$ vs. L$3$, all other tests lead to the same decision.  The second observation is that using more sampling locations does help improve the robustness of the decision, but, of course, at a higher measurement cost.  We note that the decision process based on twelve and fifteen sampling locations reached an opposite conclusion in the variance test of L$2$ vs. L$3$.  Apparently when the number of sampling locations increases, the initial difference between the variance curves is reduced, to the degree that it cannot reject the null hypothesis. Please see the variance curves on the two stages shown in Figure \ref{fig:varcurves}. 

Through this analysis, our recommendation is for the operator to use a slightly larger number of sampling locations if affordable. But in consideration of economic operation, nine sampling locations are deemed an acceptable practice, especially considering that using nine sampling locations yields a conservative decision, which is to continue polishing, rather than stopping prematurely. We also confirm that when analyzing this new set of data with more sampling locations, we did not come across any violations or contradictions of the messages obtained through the previous nine-location experiments.

\subsubsection{Sensitivity on $\tau$}\label{sec:sentau}
 Our default choice is $\tau=25\%$.  Intuitively, this $\tau$ choice says that the top quarter is the upper tail, the bottom quarter is the lower tail, and the $50\%$ in between is the  middle body.  We consider this is a reasonable choice, allocating a sufficiently large portion to cover the two tails.  For both $9$ and $15$ sampling locations, we here conducted a sensitivity analysis by varying the tail portion from a $5\%$ to $33\%$, i.e., by comparing outcomes of six different lengths of tails $\tau=5\%$, $10\%$, $15\%$, $20\%$, $25\%$ and $33\%$, respectively. The outcomes of the mean tests are listed in Table \ref{tab:senanal}.  We did not notice any significant difference in the $p$-values, and certainly no change in the detection outcomes associated with any one of the hypotheses.

\renewcommand{\arraystretch}{1.2}
\begin{table}[h!]
\begin{center}
\caption{The sensitivity of $p$-value with respect to $\tau$. }
\label{tab:senanal}
\footnotesize
\begin{tabular}{ c | c | c | c | c | c | c | c }
\hline \hline
&  & \multicolumn{6}{c}{$p$-value of mean test for upper tail} \\

\cline{3-8}
\# of Locations & Stages  & $\tau=5\%$ & $\tau=10\%$  & $\tau=15\%$ & $\tau=20\%$ & $\tau=25\%$ &$\tau=33\%$ \\
\hline
\multirow{4}{*}{$9$} & L$1$ vs. L$2$ & $0.4081$  & $0.4953$ & $0.6051$  & $0.7159$ &$0.9584$ & $0.8429$ \\
%\hline
 & L$2$ vs. L$3$ & $0.3789$ & $0.3794$  & $0.3864$ & $0.3807$ &$0.3627$ &$0.3794$ \\
%\hline
& L$3$ vs. L$4$ & $0.0001$  & $0.0003$  & $0.0003$  & $0.0002$ &$0.0004$ &$0.0004$ \\
%\hline
& L$4$ vs. L$5$ & $1$ & $1$  & $1$ & $1$ &$1$ & $1$ \\
\hline
\multirow{4}{*}{$15$} &  L$1$ vs. L$2$ &  $0.1319$ & $0.1940$  & $0.2927$  & $0.3994$ &$0.7826$ & $0.5333$ \\
& L$2$ vs. L$3$ &  $0.3629$ & $0.3605$  & $0.3590$ & $0.3518$ &$0.3318$ & $0.3473$ \\
& L$3$ vs. L$4$ & $0.0068$ & $0.0067$  & $0.0063$ & $0.0056$ &$0.0057$ & $0.0055$\\
& L$4$ vs. L$5$ & $1$ & $1$  & $1$ & $1$ &$1$ & $1$\\
\hline
&  & \multicolumn{6}{c}{$p$-value of mean test for lower tail} \\

\cline{3-8}
\# of Locations & Stages  & $\tau=5\%$ & $\tau=10\%$  & $\tau=15\%$ & $\tau=20\%$ & $\tau=25\%$ &$\tau=33\%$ \\
 \hline
 \multirow{4}{*}{$9$} & L$1$ vs. L$2$ & $0.9311$  & $0.9274$ & $0.9265$  & $0.9241$ &$0.9205$ & $0.9249$ \\
%\hline
 & L$2$ vs. L$3$ & $0.0735$ & $0.0639$  & $0.0575$ & $0.0510$ &$0.0337$ &$0.0447$ \\
%\hline
& L$3$ vs. L$4$ & $2\times10^{-5}$  & $2\times10^{-5}$  & $2\times10^{-5}$  & $2\times 10^{-5}$ &$2\times 10^{-5}$ &$2\times 10^{-5}$ \\
%\hline
& L$4$ vs. L$5$ & $1$ & $1$  & $1$ & $1$ &$1$ & $1$ \\
\hline
\multirow{4}{*}{$15$} &  L$1$ vs. L$2$ &  $0.9485$ & $0.9455$  & $0.9454$  & $0.9462$ &$0.9384$ & $0.9414$ \\
& L$2$ vs. L$3$ &  $0.3636$ & $0.3615$  & $0.3606$ & $0.3536$ &$0.3327$ & $0.3497$ \\
& L$3$ vs. L$4$ & $2\times10^{-5}$ & $2\times10^{-5}$  & $2\times10^{-5}$ & $2\times 10^{-5}$ &$2\times 10^{-5}$ & $2\times 10^{-5}$\\
& L$4$ vs. L$5$ & $1$ & $1$  & $1$ & $1$ &$1$ & $1$\\
\hline
\hline
\end{tabular}
\end{center}
\end{table}

\subsection{Method Regression Test on a Flat Surface}\label{sec:regressionexp}
As we explain earlier about a main advantage of the hypothesis test based method, it does not require geometry characteristics.  So it is certainly applicable to detecting surface quality changes on a flat surface as well. We take the data from the flat surface polishing experiment published in \cite{jin2020gaussian} and test the proposed method on it. For the details on that polishing experiment, please refer to \cite{jin2020gaussian}. Table \ref{tab:flat} summarizes the test results.

\renewcommand{\arraystretch}{1.2}
\begin{table}[h!]
\begin{center}
\caption{The hypothesis tests (HT) action suggestion versus Gaussian process based decision rule (GPBD) guided action suggestion (by \cite{jin2020gaussian}) are compared and illustrated. The confidence level $\alpha$ chosen for controlling the family of two mean tests and one variance test is $0.1$. For each of the three tests, that can be considered as independent with one another, the confidence level is $0.1/3$. }
\label{tab:flat}
\footnotesize
\begin{tabular}{  c |  c | c | c || c || c | c | c  | c}
\hline \hline
	&	 &  & Achieving &GPBD guided & Mean test & Mean test  & Variance & HT guided\\
	&	 &Pad & $\overline{Sa} $ & action &upper tail & lower tail &  test &  action \\
	From & To  & (grit) & ($\mu m$) &  suggestion & $p$-value & $p$-value & $p$-value &   suggestion\\
\hline
Stage $0$ & Stage $1$  &$800$ & $11.405$ & Continue & $2\times10^{-5}$ & $0.002$ & $0.381$ & Continue \\
Stage $1$ & Stage $2$  &$800$ & $6.510$& Continue & $0.025$ &$0.110$ &$0.793$ & Continue \\
Stage $2$ & Stage $3$  &$800$ &$1.398$ &Continue & $0.003$ &$0.001$ & $7\times10^{-5}$ & Continue\\
Stage  $3$ & Stage $4$  &$800$ &$0.304$ &Continue &$0.011$ &$0.058$ &$0.014$ & Continue\\
Stage $4$ & Stage $5$  &$800$ &$0.118$ &Continue &$0.036$ &$0.128$ &$0.009$  &Continue\\
Stage $5$ & Stage $6$  &$800$ &$0.153$ &Continue &$0.259$ &$1$ &$0.067$ & Change\\
Stage $6$ & Stage $7$  &$800$ &$0.135$ &Continue &$2\times10^{-5}$ &$0.969$ &$4\times10^{-5}$ & Continue\\
Stage $7$ & Stage $8$  &$800$ &$0.144$ &Continue &$0.883$ &$1$ &$0.996$ & Change\\
Stage $8$ & Stage $9$  &$800$ &$0.163$ &Continue &$1$ &$1$ &$0.859$ & Change \\
Stage $9$ & Stage $10$ &$800$ &$0.141$ &Continue &$0.638$ &$2\times10^{-5}$ &$9\times10^{-6}$ & Continue\\
Stage $10$ & Stage $11$ &$800$ &$0.174$ & \textbf{Change} &$1$ &$1$ &$1$ & \textbf{Change}\\
Stage $11$ & Stage $12$ &$1200$ &$0.180$ &Continue &$0.992$ &$0.945$ &$0.385$ & Change\\
Stage $12$ & Stage $13$ &$1200$ &$0.094$ &Continue &$0.001$ & $2\times10^{-5}$&$0.018$ &Continue\\
Stage $13$ & Stage $14$ &$1200$ &$0.169$ &Continue &$1$ &$1$ &$1$ & Change\\
Stage $14$ & Stage $15$ &$1200 $&$0.171$ &Continue &$1$ &$0.972$ &$0.003$ & Continue\\
Stage $15$ & Stage $16$ &$1200$ &$0.165$ &Continue &$0.101$ &$0.104$ &$0.491$ & Change\\
Stage $16$ & Stage $17$ &$1200$ &$0.137$ &Continue &$0.031$ &$8\times10^{-5}$ &$0.668$ &Continue\\
Stage $17$ & Stage $18$ &$1200$ &$0.207$ &\textbf{Change} &$1$ &$1$ &$0.989$ &\textbf{Change}\\
Stage $18$ & Stage $19$ &$1200$ &$0.128$ &Continue &$0.001$ &$2\times10^{-5}$ &$4\times10^{-5}$ &Continue\\
Stage $19$ & Stage $20$ &$1200$ &$0.120$ &Continue &$0.056$ &$0.043$ &$8\times10^{-5}$ &Continue\\
Stage $20$ & Stage $21$ &$1200$ &$0.116$ &Continue &$0.344 $ &$0.405$ &$0.948$ &Change\\
Stage $21$ & Stage $22$ &$1200$ &$0.140$ &\textbf{Change} &$1$ &$1$ &$0.868$ &\textbf{Change}\\
Stage $22$ & Stage $23$ &*MC &$0.061$ &Continue &$2\times10^{-5}$ &$2\times10^{-5}$ &$0.021$ &Continue\\
Stage $23$ & Stage $24$ &*MC &$0.053$ &Continue &$0.124$ &$0.046$ &$0.343$ &Change\\
Stage $24$ & Stage $25$ &*MC &$0.054$ &\textbf{Change} &$0.453$ &$0.502$ &$0.711$ &\textbf{Change} \\
\hline
\hline
\multicolumn{9}{r}{* MC stands for microcloth.}\\
\end{tabular}
\end{center}
\end{table}

The right half of Table \ref{tab:flat} presents the $p$-values of the three hypothesis tests: the upper tail mean test,  the lower tail mean test, and the variance test. Consistent with studies presented in the two preceding subsections, the $p$-value cut-off used is $0.1$ at the overall decision level, or $0.033$ at the individual test level. As long as any of the three hypothesis tests is rejected, a change in surface quality is considered being detected and the action suggested is to keep polishing with the current tool. Otherwise, no surface change is detected and the action suggestion is to either clean or change the current polishing tool, or stop altogether.

The column ``GPBD guided action suggestion'' shows the suggestions given by the Gaussian process (GP)-based decision rule proposed by \cite{jin2020gaussian}. The GP-guided decision process is to compare the similarity between sampling locations, quantified by the scale parameter in the GP model, between Stage $t-1$ with Stage $t$. \cite{jin2020gaussian} deem the across-stage similarity a good proxy to inform about the roughness of the surface. The decision is then made by comparing the change in the scale parameter in the Gaussian process models associated with the two consecutive stages.

By comparing the decision outcomes from the two methods, we notice that every time when GPBD suggests a change, HT-based method also suggests the same. On the other hand, HT-based method suggests more changes, and more importantly, earlier change points.  Should the HT-based method be followed, the first change action would have been at Stage 6, five stages earlier than that suggested by GPBD. This suggestion means that one ought to change the polishing tool from its initial $800$-grit pad to a finer $1,200$-grit pad.  By looking at the value of $\overline{Sa}$, we can see the merit of this new suggestion. From Stage 6 through Stage 11, $\overline{Sa}$ is fluctuated between $0.135 \mu m$ and $0.174 \mu m$ with no clear sign of reducing. It appears that using the $800$-grit was effective to bring down the surface roughness from a very rough raw surface to the level of $0.150 \mu m$, namely $150 nm$, but was ineffectiveness to make further inroads. This may be the result of the inherent capability of the $800$-grit. Had one changed the pad after Stage 6, the next two changes suggested by the HT-based method may not necessarily happen anymore, because once a new pad is used, the product surface would react differently and the change in the surface roughness usually follows a different course.

From the method design point of view, we can also see that the HT-based decision process presents certain advantages over the GPBD-guided decisions. As explained above, GPBD characterizes the spatial correlation between sampling locations; the stronger the correlation, the smoother the surface. This is reasonable, but the information in the roughness is aggregated into a scalar correlation parameter. \cite{jin2020gaussian} demonstrated that this scalar correlation parameter is still more informative than the median $Sa$ value. Yet, in the process of condensing the information associated with BACs into scale parameters, certain fine granularity of information could get lost. By contrast, the HT-based method detects a surface quality change by comparing two set of curves without the data compression steps, so it appears more robust and preferred.

\section{Concluding Remarks}\label{sec:conclusion}
We proposed in this paper a method that does not require geometry characterization for enabling endpoint decisions in surface finishing processes of precision manufacturing products. We illustrate its applicability to spherical beads polishing/lapping processes.  The method detects the surface quality lack-of-change point by comparing the curve clusters of two consecutive polishing stages. If none of the three situations happens---the lowered upper tail, the raised lower tail or the reduced variance, the surface quality lack-of-change point is detected. Then, it calls a need of polishing action change into one's attention. 

There are a number of unique features of the method worth highlighting. Compared with GP-based or other model-based methods, the proposed decision method does not need to specify an underlying model and rely on strong assumptions. The proposed method is built upon statistical testing of nonparametric quantile curves, which are a direct result of sorting the measurements of surface roughness. Such quantile curves, known as the bearing area curves in engineering, are readily available in many different kinds of finishing processes.  This explains that the method can be applied to both lapping and polishing processes, both spherical and flat surfaces, without changes in its underlying procedure. The method is able to test the relationship of quantile curves for a specified region, or even, a union of disconnected sub-regions, should prior physical knowledge advise such choices. Compared with the surface roughness measure like $Sa$, the proposed decision method not only reveals the detailed nature of a change in surface roughness, but also informs the statistical significance of such change.

Adopting the proposed decision process leads to earlier stopping or tool-changing actions, which could save a lot of time, energy, or material removal, without sacrificing the final surface finishing quality.  More than that, shortening the final long stretch of polishing has the added benefit of reducing the risk that the polishing products may be cracked, scratched, or otherwise damaged.

\appendix
\section{Appendix: Type II Error.}\label{secA1}
We simulate two groups of $N$ functions to estimate the type II error of the mean test for upper tail and lower tail, while controlling the type I error to be under the nominal level, i.e., $\alpha=0.03$. The two groups of functions to be tested are simulated from a Gaussian process, with one group of functions digressing from the other group by a small perturbation. To quantify the small perturbation between two groups of functions, we use a $L^2$-distance percentage defined as:
\begin{equation}
    L^2 \% = \frac{\|\mu_1-\mu_2\|_{L^2}}{\|\mu_1\|_{L^2}}\times 100\%, 
\end{equation}
where $\mu_i$ is the mean function of the functions of group $i$. 

\begin{figure}
    \centering
    \includegraphics[width=0.5\textwidth]{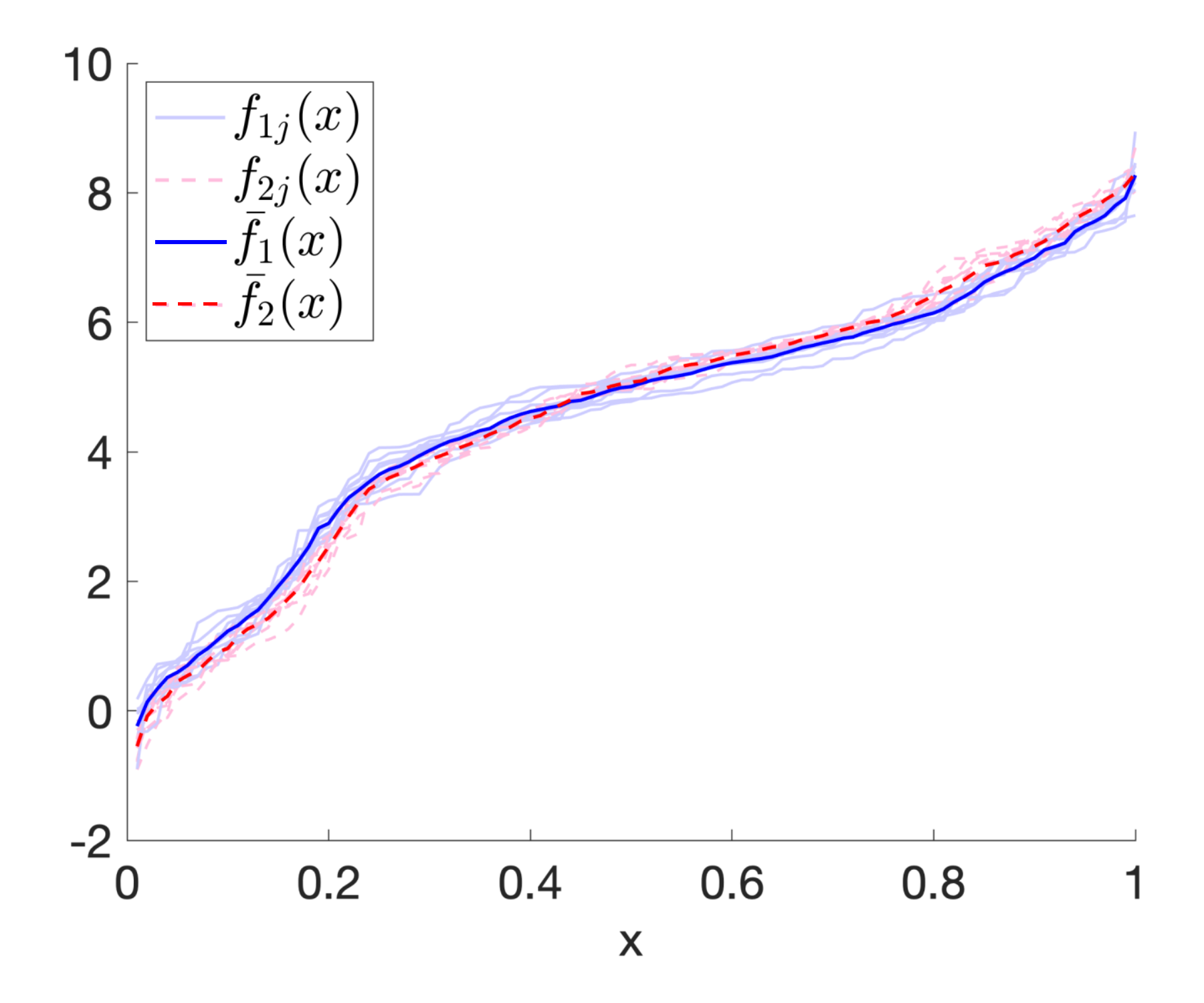}
    \caption{The two groups of nine functions, $f_{1j}(x)$ and $f_{2j}(x)$, simulated from one run, and the mean functions, $\bar{f_1}(x)$ and $\bar{f_2}(x)$. }
    \label{fig:12}
\end{figure}

To generate the two groups of $N$ functions, we randomly sample one set of input points, $x\in[0,1]$, as the pointwise test requires that two groups of functions have to be evaluated at the same input points. The two groups of functions are generated from the model described as: $f_{ij}(x)=z(x)+\epsilon_{ij}$, $i\in\{1,2\}$, $j\in\{1,\cdots,N\}$; $z(x)\sim GP(\mathbf{0},k(x,x'))$; $\epsilon_{ij}\sim N(0,\sigma_\epsilon^2)$ and $\sigma_{\epsilon}=0.5$. The covariance function $k(x,x')$ is a squared exponential kernel function with the form: $k(x,x')=\sigma_f^2 \exp(-0.5[(x-x')/\theta]^2)$. We set $\sigma_f=5$ and $\theta=0.2$. To make the second group of functions that deviates from the first group, we add a perturbation $\delta(x)$ to $f_{2j}(x)$, $j\in\{1,\cdots,N\}$. The perturbation function $\delta(x)$ is created as:

\begin{equation}
\delta(x) = 
    \begin{cases} 
      -\frac{1}{3}\sin\left(\pi\left(\frac{x-0.2}{0.8-0.2}\right)\right), & x\leq 0.25, \\
      0, & 0.25< x< 0.75, \\
      \frac{1}{3}\sin\left(\pi\left(\frac{x-0.2}{0.8-0.2}\right)\right), &  x \geq 0.75.
  \end{cases}
\end{equation}
Thus, $f_{1j}(x) = z(x)+\epsilon_{1j}$ and $f_{2j}(x) = z(x) + \delta(x)+\epsilon_{2j}$, $j=1,\cdots,N$. We generate $N = 6,9,12,15$ functions for each of the two groups by sampling $\epsilon_{ij}$. Run our proposed hypothesis test for the mean tails and repeat the process for $1000$ runs. The functions $f_{1j}(x)$ and $f_{2j}(x)$ generated for one run and their mean functions $\bar{f_1}(x)$ and $\bar{f_2}(x)$ are shown in Figure \ref{fig:12}. 

The average $L^2$ distance percentage over between two groups of functions over $1000$ runs is $3.52\%$. The type II error of the hypothesis test on the simulation data, with the type I error controlled under $0.03$, varies with the number of curves, $N$. Table \ref{tab:typeII} shows the estimated type II error and the average $L^2$ distance percentage corresponding to each $N$.

\begin{table}[htb]
\begin{center}
\caption{Estimated type II errors for the simulated $N$ curves, with type I error under control.}
\label{tab:typeII}
\footnotesize
\begin{tabular}{  c | c | c | c }
\hline 
\hline
\multirow{2}{*}{Number of curves, $N$} & \multirow{2}{*}{Average $L^2$ distance percentage} & \multicolumn{2}{c}{Type II error}  \\
\cline{3-4}
& & Mean test for upper tail & Mean test for lower tail\\
\hline
$6$ & $3.61\%$ &$0.224$ & $0.383$ \\
$9$ & $3.54\%$ &$0.156$ & $0.272$ \\
$12$ & $3.49\%$ & $0.112$ & $0.229$\\
$15$ & $3.43\%$ & $0.091$ & $0.174$ \\
\hline
\hline
\end{tabular}
\end{center}
\end{table}

The type II error decreases as the number of curves of each group increases, while the distance between two groups of curves stays more or less the same. That is consistent with the commonsense of a large sample size increasing the power of a hypothesis test. 

However, in practice it is quite consumable and, sometimes, infeasible to draw many samples. As in our polishing experiment, we are only able to image nine to $15$ locations on the spherical surface of peppercorn-sized bead. Although the $0.1$ to $0.2$ type II error rate may lead to an early change of polishing actions, e.g., changing pad or stopping polishing, this decision making will not necessarily worsen the surface finishing quality, considering the small difference between the two groups, but may save the polishing effort, prevent the material from being excessively removed and reduce the possibility of creating defects on the surfaces.

\if0\blind{
\section*{Acknowledgements}
The authors acknowledge the generous support from National Science Foundation under grant no. IIS-1849085, Department of Energy grant no. DE-AC52-07NA27344 through Subcontract no. B646055, and Texas A\&M Office of President's X-grant Program.
% The authors acknowledge the generous support from National Science Foundation under grant no. ***-*******, Department of Energy grant no. **-****-********* through Subcontract no. *******, and another funding agent (eliminated for double-blinded review purpose).
} \fi

\bibliographystyle{chicago}
\spacingset{1}
\bibliography{HTGP}
	
\end{document}